\documentclass[format=acmsmall, review=false]{acmart}
\usepackage{acm-ec-26}
\usepackage[T1]{fontenc}
\usepackage{mathtools}
\usepackage{bm}
\usepackage{booktabs}  
\usepackage{float}     
\usepackage{tikz}
\usepackage{forest}
\usepackage{bbm}
\usetikzlibrary{arrows.meta}
\usepackage[ruled]{algorithm2e} 

\SetAlFnt{\small}
\SetAlCapFnt{\small}
\SetAlCapNameFnt{\small}
\SetAlCapHSkip{0pt}
\IncMargin{-\parindent}
\newcommand{\thmheadfont}{\bfseries}
\usepackage{array}   
\newcolumntype{C}{>{$}c<{$}} 
\usepackage{tabularx} 

\usepackage{amsthm}

\newcounter{exctr}
\renewcommand{\theexctr}{\arabic{exctr}}

\usepackage{microtype}
\usepackage{setspace}

\theoremstyle{plain}

\newtheorem{result}{Result}

\newtheorem{hypothesis}{Hypothesis}

\makeatletter
\renewcommand{\thmhead}[3]{%
	\thmheadfont{#1\ #2}%
	\thm@notefont{: [#3]}%
}
\makeatother

\newcommand{\rk}{\textup{rk}}

\setcitestyle{authoryear}

\title{Experimental School Choice with Parents}

\author{Mikhail Freer}
\affiliation{%
	\institution{University of Essex}
	\country{UK}
}

\author{Thilo Klein}
\affiliation{%
	\institution{Pforzheim University and ZEW}
	\country{Germany}
}

\author{Josu\'e Ortega}
\orcid{0000-0002-5080-2058}
\affiliation{%
	\institution{Queen's University Belfast}
	\country{UK}
}
\begin{abstract}
	\vspace*{2em}
	We conduct the first laboratory school choice experiment in which parents -- the relevant decision makers in the field -- are the experimental subjects.
	We compare Deferred Acceptance (DA) with two manipulable but potentially more efficient alternatives: Efficiency-Adjusted Deferred Acceptance (EADA) and the Rank-Minimizing mechanism (RM).
	
	We find that all mechanisms are frequently manipulated, with no significant differences in truth-telling rates. Parents and students manipulate at similar rates, supporting the external validity of student-based experiments, though students make significantly more obvious errors, suggesting parents' deviations are more deliberate. Despite widespread manipulation, the predicted welfare–stability tradeoff largely survives: DA never produces Pareto-efficient allocations yet generates little justified envy;  whereas RM delivers substantial efficiency gains at a meaningful stability cost. EADA occupies a middle ground: its efficiency gains over DA are modest and imprecisely estimated yet double justified envy. 
	Higher cognitive ability is associated with more deviations, and under EADA with worse outcomes.
	While DA does not induce truth-telling, it is the only mechanism in which manipulation never pays off and rarely changes outcomes.
	
	
\end{abstract}

\begin{document}
	\begin{titlepage} 
		\maketitle \makeatletter \gdef\@ACM@checkaffil{} \makeatother
		\thispagestyle{empty} 
		
	\end{titlepage}

	\setlength{\parskip}{6pt plus 2pt minus 1pt}
	
	\onehalfspacing
	\section{Introduction}

	The redesign of student assignment mechanisms is one of the most visible contributions of economic theory to public policy.
	Beginning with \citet{abdulkadirouglu2003}, economists showed that school admissions procedures in several American cities induced strategic behavior, giving an advantage to families who better understood the allocation process.
	A central policy response has been the adoption of strategy-proof mechanisms such as the student-proposing Deferred Acceptance (DA) algorithm \citep{gale1962}.
	DA achieves two important objectives.
	First, truthful reporting is a weakly dominant strategy, so sophisticated families cannot gain an advantage through strategic preference manipulation.
	Second, under truthful reporting, DA produces assignments that respect schools' priorities: put differently, it does not generate any justified envy.
	These advantages have driven DA's widespread adoption in school admissions around the world.
	
	As DA became the policy benchmark, attention naturally turned to the question of whether parents in practice behave in the truth-telling manner that theory prescribes.
	Although DA makes truth-telling a weakly dominant strategy, observed behavior may still depart from truthful reporting.
	Such departures can generate justified envy relative to agents' true preferences, even though DA's outcome is stable with respect to the submitted rankings.
	To study behavior under controlled conditions, economists have conducted laboratory experiments in which subjects rank artificially created schools, with monetary incentives tied to their assignment.
	These experiments allow researchers to compare subjects' induced true preferences with their submitted rankings, and to measure the impact of any manipulations on the efficiency and justified envy of the corresponding assignments.
	The large experimental school choice literature, beginning with \citet{chen2006school}, has consistently found that DA experiences substantially less manipulation than the Boston mechanism.
	These experimental findings also influenced policy decisions. In the words of \citet{calsamiglia2011comment}, \emph{``it played an important role to convince the Boston school district authorities to replace the previous mechanism by one of the other mechanisms.''}
	
	Nonetheless, despite being extensively replicated under a number of robust environments in the lab, the external validity of lab-based school choice remains unclear, as findings about students' behavior in school allocation mechanisms may or may not generalize to those of actual parents.
	Differences in cognitive sophistication, among several other characteristics in which these two populations may differ, may be especially relevant for strategic behavior in matching mechanisms \citep{basteck2018cognitive}.
	Because parents and undergraduates may differ systematically along these dimensions, evidence from student-subject experiments need not reflect parental behavior.
	As \citet{levitt2007generalizability} note, ``\emph{there is no question more fundamental to experimental economics than understanding whether laboratory results generalize to naturally occurring environments}''.
	Yet, despite influencing education policy worldwide, the school choice literature has not, to the best of our knowledge, directly tested this generalization from undergraduates to parents in a controlled experiment.\footnote{There is, however, a large literature beyond school choice that analyzes the representativeness of standard laboratory subjects. We discuss it in Section~\ref{sec:literature}.}
	
	Our main contribution is to conduct the first laboratory experiment on school choice mechanisms in which parents, rather than students, serve as the experimental subjects.
	Using this field-in-the-lab setting directly addresses the external validity gap by observing how the relevant decision makers behave when confronted with different assignment rules.
	Studying parents allows us to examine behavior under markedly different demographic and decision-making characteristics than those of undergraduates.
	By comparing parents' reports with those of students under otherwise identical experimental conditions, we provide the first direct evidence on whether the behavioral patterns documented in student-based studies extend to the population for whom these mechanisms are ultimately designed.
	
	A second contribution is to experimentally evaluate two alternative and relatively novel mechanisms alongside DA: Efficiency-Adjusted Deferred Acceptance (EADA, \citealt{kesten2010school}) and, for the first time in any experimental setting, the Rank-Minimizing mechanism (RM, \citealt{featherstone2020rank}).
	These mechanisms represent different points on the efficiency-stability frontier.
	EADA is designed to deliver Pareto improvements over the DA outcome under truthful reporting, limiting justified envy.
	RM is also Pareto-efficient, but goes further by minimizing the sum of assigned ranks, at the cost of disregarding priorities entirely.
	RM therefore serves as a benchmark for rank-efficiency: it quantifies how much welfare can be achieved if one is willing to give up on schools' priorities.
	Both mechanisms, in particular EADA, have received substantial interest from theorists because, although manipulable, the incentives to deviate are not obvious and thus they could deliver large efficiency improvements over DA \emph{if} subjects report their preferences truthfully \citep{troyan2020obvious,troyan2022non}.
	Moreover, while EADA does not fully respect priorities, it satisfies several weaker but meaningful stability notions (described in Section~\ref{sec:literature}), which have made it a prominent mechanism in the literature and a candidate for real-life implementation in Flanders.
	Our contribution is to test these mechanisms in the lab and to document whether manipulations eliminate their efficiency guarantees under truthful reporting.
	
	\paragraph{Results preview.}
	We recruit 324 parents and 216 students and assign them to one of five treatments: parents face DA, EADA, or RM. Students face DA under equal or lower stakes for comparison. We match them with their peers in groups of 18 subjects, competing for 18 seats distributed among seven schools (i.e. parents interact with parents, and students with students).
	Subjects are informed about the mechanisms and the ranking interface, tested on their mechanism understanding, and given a practice round in the ranking task. We induce preferences over schools using a convex payoff function, guaranteeing a payment of £55 if they are assigned to their top choice. The experiment is designed to mimic real-life school choice: there is substantial but incomplete information and parents only rank schools once. In particular, subjects know how many seats are available at each school, the number of competitors who rank each school as top choice, and their priority at each school. After ranking schools, we measure cognitive ability using incentivized Raven matrices, as \citet{basteck2018cognitive}, to detect differences in cognitive ability between parents and students and see how they affect truth telling and outcomes. 
	
	First and foremost, we find high manipulation rates across strategy-proof and manipulable mechanisms, and across both parents and students. Only 26\% of parents report truthfully under DA, statistically indistinguishable from EADA (29\%), RM (23\%) or students under DA (22\%). Gross misunderstanding is unlikely to explain these deviations: 88\% of subjects pass comprehension quizzes. Yet what distinguishes the mechanisms is not how often subjects manipulate but what happens when they do: in manipulable mechanisms, significantly more manipulations are consequential, and only under RM do some (6\%) lead to higher payoffs. 
	
	Second, while parents and students manipulate at roughly the same aggregate rates, providing the first direct evidence that findings from student-based experiments generalize to parents, we find that students make obvious mistakes significantly more often. An obvious mistake is not ranking the most preferred school as first choice whenever this school is a safe one (where one has a priority higher than the number of seats available). Parents rarely make such errors, suggesting their deviations are more deliberate, even if ultimately unproductive or harmful. This has a practical implication: if even parents who demonstrably understand the mechanism deviate from truthful reporting, policymakers cannot rely on strategy-proofness to induce truthful behaviour. Rather, the mechanism must protect agents from the standard yet flawed manipulation heuristics they inevitably employ, which is precisely what DA achieves in our experiment. Under EADA and RM, where consequential manipulations are significantly more frequent, significantly more subjects are harmed by their own deviations.
	
	Third, the theoretical ranking of mechanism properties predicted under truth-telling is mainly preserved despite widespread manipulation, but with a surprising twist.
	As expected, DA is the mechanism with fewest priority violations, yet it pays a very high efficiency cost: it never generates a Pareto-efficient allocation.
	RM delivers substantial efficiency gains over DA: the average assigned rank improves by nearly half a position (out of seven, 2.35 vs.\ 2.84), and the worst-off student improves by almost a full rank (5.86 vs.\ 6.80). Both represent significant improvements in an environment with seven schools, but these gains come at the cost of half of students experiencing justified envy.
	EADA's theoretical appeal, by contrast, does not translate into meaningful welfare gains: the efficiency improvement over DA is smaller (2.74 vs.\ 2.84) and the fraction of assignments that are efficient is indistinguishable from 0, yet EADA  generates justified envy for 29\% of students. Although theoretically impossible under truth-telling, we find that EADA generates an even worse worst-off placement than DA (6.94 vs.\ 6.80, the difference is not statistically significant).
	EADA thus represents a compromise between efficiency and stability that achieves neither goal in our experiment, paying a stability cost while buying little efficiency.
	
	Finally, our analysis of cognitive ability yields a further, unexpected result. Higher Raven scores are associated with less truthful reporting across all mechanisms. Yet the consequences of these deviations vary sharply by mechanism. Under DA, higher-ability subjects achieve slightly better assignments, whereas under EADA the point estimates reverse: higher-ability subjects achieve worse outcomes than their lower-ability counterparts, consistent with a sophistication trap in which the subjects most inclined to strategize are precisely those most harmed by doing so under a mechanism whose profitable deviations are non-obvious. This pattern is directionally robust across ability measures, though its statistical strength varies.
	
	Taken together, our results suggest that DA protects families through two distinct channels. Strategy-proofness ensures that no deviation can improve one's assignment. But, in our data, DA also exhibits a second valuable property: most deviations do not alter the assignment at all. It is this combination, and not strategy-proofness alone, that shields families, including the most strategically inclined parents, from the costs of their own deviations. 
	\section{Related Literature}
	\label{sec:literature}
	
	\paragraph{Experimental School Choice.}
	The experimental literature on school choice begins with the seminal study of \citet[][henceforth CS06]{chen2006school}.
	In CS06, $36$ students competed for $36$ school seats distributed across $7$ schools, and assignments were generated by Deferred Acceptance (DA), the Boston mechanism (BOS), or Top Trading Cycles (TTC).
	Subjects were asked to submit a complete ranking of schools, and earnings were tied to the rank of the assigned school.
	To mimic key features of real-world school choice, the game was played only once and each subject faced only one mechanism.
	CS06 studied two preference environments: one with independent preferences and one with a realistic correlation pattern.
	For each mechanism and each environment, they ran two sessions.
	Subjects knew their own payoffs, the set of schools, and capacities, but had no information about other subjects' preferences.
	To conduct inference with a small number of sessions, CS06 introduced a {recombinant} estimation procedure, which permutes subjects' submitted rankings across sessions within the same treatment to construct counterfactual groups and approximate the distribution of outcomes.
	
	A large literature has since replicated and extended CS06 along many dimensions: restricting list lengths, varying information and advice, eliciting or correlating behavior with characteristics such as risk aversion and cognitive ability, introducing experience or repeated play, testing additional mechanisms, changing the timing of preference revelation, scaling the market size, and adapting the environment to object allocation rather than monetary payoffs.
	We summarize this literature in Table~\ref{tab:literature1}; see \citet{hakimov2021experiments} for a detailed overview.
	\begin{table}[t]
		\caption{Experimental School Choice Literature.}
		\label{tab:literature1}
		\resizebox{\linewidth}{!}{%
			\begin{tabular}{llr}
				\toprule
				Study & Mechanisms & Participants \\
				\midrule
				\citet{chen2006school} & DA, TTC, BOS & 432 students \\
				\citet{pais2008school} & DA, TTC, BOS & 435 students \\
				\citet{calsamiglia2010constrained} & DA, TTC, BOS & 864 students \\
				\citet{klijn2013preference} & DA, BOS & 218 students \\
				\citet{zhu2015experience} & DA & 216 students \\
				\citet{castillo2016truncation} & DA & 92 students \\
				\citet{chen2016school} & DA, TTC, BOS & 175 students \\
				\citet{klijn2016affirmative} & DA, TTC& 175 students\\
				\citet{featherstone2016boston} & DA, BOS & 55 students \\
				\citet{ding2017matching}& DA, BOS& 510 students\\
				\citet{guillen2017not} & TTC & 202 students \\
				\citet{basteck2018cognitive} & DA, BOS & 192 students \\
				\citet{chen2018matching}&DA, BOS& 736 students\\
				\citet{kawagoe2018skipping} & DA & 90 students \\
				\citet{chen2019chinese} & IA, PA, DA & 360 students \\
				\citet{ding2019learning}& DA, BOS & 375 students\\
				\citet{dur2019secure} & DA, BOS, sBOS & 81 students \\				
				\citet{klijn2019static}&sDA,dDA&192 students\\
				\citet{bo2020iterative} & DA, IDAM, IDAM-NC & 288 students \\
				\citet{klijn2020improving} &DA & 96 students\\
				\citet{basteck2021lotteries}&DA, BOS & 384 students\\
				\citet{castillo2021strategic} & DA-TT, DA-TR & 120 students \\
				\citet{dur2021sequential} & DA, BOS & 48 students \\
				\citet{guillen2021strategy} & DA, TTC, RDA, RTTC & 209 students \\
				\citet{cho2022tie}&DA, SIC, CADA&639 students\\
				\citet{koutout2021mechanism} & DA, BOS & 288 students \\
				\citet{afacan2022parallel} & DA & 92 students \\
				\citet{stephenson2022assignment} & DA, TTC, BOS & 432 students \\
				\citet{cerrone2022school} & DA, EADA & 500 students \\
				\citet{hu2024cognitive}&DA&176 students\\
				\citet{hakimov2025} & sDA, dDA & 186 students \\
				\midrule
				\textbf{This paper} & \textbf{DA, EADA, RM} & \textbf{324 parents (+ 216 students)} \\
				\bottomrule
		\end{tabular}}
	\end{table}

	Two robust findings emerge.
	First, BOS is manipulated substantially more than DA.\footnote{Some studies also find that DA yields more Pareto-efficient outcomes than BOS, though this conclusion can depend on the recombination procedure and the number of recombinations used \citep{chen2011corrigendum,calsamiglia2011comment}.}
	Second, all mechanisms are manipulated in the lab, including strategy-proof ones.
	For example, the truth-telling rate under DA in CS06 ranges from 55\% to 72\% depending on the environment, and subsequent studies document rates that vary widely across designs: from as low as 16\% \citep{guillen2021strategy} to above 90\% \citep{pais2008school,featherstone2016boston}.
	
	Within this literature, the closest study to ours is \citet[][henceforth CHK24]{cerrone2022school}, which is the only previous experiment to test EADA (to our knowledge, no experiment has studied RM).
	CHK24 compare DA with three variants of EADA that differ in the consent structure (opt-in, opt-out, and forced consent).
	Their design features repeated interaction: subjects are organized into session groups and repeatedly re-matched over 20 periods, while facing the same mechanism throughout.
	Subjects' preferences remain constant over time, and participants have perfect information about the environment; they also receive no recommendation to report truthfully.
	Payoffs are expressed in points (25, 18, 12, 7, 3 for first through fifth choice), and earnings are based on two randomly selected rounds (with 4 points equal to €1), plus a show-up fee; the average payment is €14.5.
	
	CHK24 document substantial efficiency gains from EADA relative to DA.
	Depending on the consent structure, EADA yields a Pareto-efficient assignment in 44\% to 80\% of cases, whereas only 6\% of DA assignments are Pareto-efficient in their design.
	They also find higher truth-telling under EADA than under DA (67\% versus 44\%).
	Finally, the fraction of stable assignments is statistically similar across DA and EADA variants and ranges between 68\% and 81\%.
	
	Our study complements CHK24 along three dimensions.
	First, we focus on a one-shot environment with incomplete information, following the CS06 tradition, which is intended to mimic the informational constraints present in real-world school choice settings at the cost of not being able to measure learning over time.
	Second, we directly address external validity by recruiting parents as subjects and comparing their behavior to students under otherwise identical conditions.
	Third, we provide the first experimental evidence on RM, which serves as a benchmark for the rank-efficiency gains available, allowing us to understand the magnitude of EADA's efficiency gains against the first-best.
	
	\paragraph{External Validity and Subject Pools.}
	
	The external validity of laboratory experiments has been debated since the inception of experimental economics \citep{smith1980relevance}.
	A central concern is whether lab evidence obtained from standard subject pools -- most commonly undergraduates -- generalizes to naturally occurring environments \citep{levitt2007generalizability,camerer2011promise}.
	One strand of the debate emphasizes that monetary stakes in the lab are modest relative to many real-world decisions, potentially affecting effort and strategic reasoning.
	Another strand argues that laboratory environments are nonetheless valuable for isolating mechanisms and identifying robust qualitative patterns, even if magnitudes differ in the field.
	
	A large empirical literature compares behavior across subject pools, including students, non-students, and online samples (e.g.\ MTurk), and studies how selection and demographics affect experimental findings \citep{andersen2010preference,anderson2013self,belot2015comprehensive,cappelen2015social,frechette2015laboratory,frigau2019field,henrich2001search,snowberg2021testing}.
	We build on this literature by focusing on a particularly policy-relevant subject pool -- parents making school-choice-like decisions -- and by examining whether the behavioral regularities documented in student-subject school choice experiments extend to this population.
	
	\paragraph{EADA and RM Properties.} Over the past decade, EADA's properties and implementation have been extensively studied \citep{bando2014existence, tang2014new, dur2019school, dur2020you, troyan2020obvious,  tang2021weak, dougan2021minimally, chen2023regret,dogan2023existence,afacan2025improving,shirakawasimple,ortega2025pareto,ortega2024identifying}, demonstrating that it is possible to achieve an efficient improvement over DA while maintaining relatively low instability and manipulability. Interestingly, different normative frameworks on how to relax stability, such as legality and priority-neutrality, converge to EADA's outcome as the student-optimal matching in these frameworks \citep{ehlers2020legal,reny2022efficient}.
	
	Much less is known about RM. Proposed by \citet{featherstone2020rank} (but widely studied in operations research, see \citet{aldous2001zeta} and \citet{parviainen_2004}), it has been shown to produce significantly better rank distributions than other Pareto-efficient mechanisms, and surprisingly, even less justified envy than efficient mechanisms that take priorities into account, like TTC \citep{nikzad2022rank,ortega2023cost}. Recent work has investigated how subjects can manipulate this mechanism, but such analysis has not been conducted empirically \citep{troyan2022non,tasnim2024strategic,okumura2024strategic,bando2024characterization}.
	
	\section{Model}
	\label{sec:model}
	
	A school choice problem $P$ consists of: (i) a finite set of students $I$; (ii) a finite set of schools $S$, where each school $s \in S$ has capacity $q_s \in \mathbb{N}$; (iii) for each student $i \in I$, a strict preference order $\succ_i$ over schools; and (iv) for each school $s \in S$, a strict priority order $\triangleright_s$ over students.
	A \emph{matching} is a function $\mu:I \to S$ such that for every school $s \in S$,
	$\left|\mu^{-1}(s)\right| \leq q_s$.
	We write $\mu(i)$ for the school assigned to student $i$.
	For each student $i$, let $\rk_i(s)$ denote the rank of school $s$ in $i$'s preference order $\succ_i$, with $\rk_i(s)=1$ for $i$'s top choice and larger values corresponding to less-preferred schools. We write $\rk_i(\mu(i))$ for the rank of $i$'s assignment under $\mu$.

	A matching $\mu$ \emph{Pareto dominates} a matching $\nu$ if $\rk_i(\mu(i)) \le \rk_i(\nu(i))$ for all $i \in I$ and $\rk_j(\mu(j)) < \rk_j(\nu(j))$ for some $j \in I$.
	A matching is \emph{Pareto-efficient} if it is not Pareto dominated.
	A stronger efficiency notion is \emph{rank efficiency}: $\mu$ is rank efficient if there is no matching $\nu$ such that
	$\sum_{i \in I} \rk_i(\nu(i)) \;<\; \sum_{i \in I} \rk_i(\mu(i))$.\footnote{Rank efficiency implies Pareto-efficiency, but the converse need not hold.}
	
	A matching $\mu$ is \emph{blocked} by a student-school pair $(i,s)$ if (i) $i$ prefers $s$ to her assignment, i.e.\ $s \succ_i \mu(i)$, and (ii) either $s$ has an unfilled seat under $\mu$ or there exists a student $j$ with $\mu(j)=s$ such that $i \triangleright_s j$.
	In that case, we say that $i$ experiences \emph{justified envy}.
	A matching with no blocking pair (equivalently, with no student experiencing justified envy) is called \emph{stable}.
	
	\subsection{Three Mechanisms of Interest}
	A \emph{mechanism} selects a matching for each school choice problem.
	A mechanism is \emph{Pareto-efficient} (resp.\ \emph{rank efficient}, \emph{stable}) if for every problem it selects a matching that is Pareto-efficient (resp.\ rank efficient, stable).
	A mechanism is \emph{strategy-proof} if truthful preference reporting is a weakly dominant strategy for every student in the associated preference revelation game.
	
	We consider three mechanisms: student-proposing Deferred Acceptance (DA, \citealt{gale1962}), Efficiency-Adjusted Deferred Acceptance (EADA, \citealt{kesten2010school}), and the Rank-Minimizing mechanism (RM, \citealt{featherstone2020rank}).
	We describe each mechanism and illustrate it using the example in Table~\ref{tab:example} (with one seat per school).
	
	\begin{table}[H]
		\centering
		\caption{A school choice problem with four students and four schools.}
		\label{tab:example}
		\begin{tabular}{ccccccccc}
			
			\multicolumn{4}{c}{\textbf{Preferences}} & \phantom{xxx} & \multicolumn{4}{c}{\textbf{Priorities}}\\
			\cmidrule(r){1-4}\cmidrule(l){6-9}
			$i_1$ & $i_2$ & $i_3$ & $i_4$ & & $s_1$ & $s_2$ & $s_3$ & $s_4$ \\
			\midrule
			$s_2$ & $s_2$ & $s_4$ & $s_4$ & & $i_1$ & $i_4$ & $i_2$ & $i_1$ \\
			$s_3$ & $s_4$ & $s_3$ & $s_3$ & & $i_4$ & $i_3$ & $i_1$ & $i_2$ \\
			$s_1$ & $s_3$ & $s_2$ & $s_2$ & & $i_3$ & $i_1$ & $i_4$ & $i_4$ \\
			$s_4$ & $s_1$ & $s_1$ & $s_1$ & & $i_2$ & $i_2$ & $i_3$ & $i_3$ \\
			\bottomrule
		\end{tabular}
	\end{table}
	
	\paragraph{Deferred Acceptance (DA).}
DA proceeds in rounds.
In round 1, each student applies to her top-ranked school.
Each school tentatively holds up to $q_s$ applicants with highest priority among those who have applied so far, and rejects the rest.
In each subsequent round, any rejected student applies to her next most-preferred school not yet applied to.
Each school again tentatively holds its highest-priority applicants up to capacity among all applicants it has seen so far, rejecting any others.
The process terminates when no student is rejected; tentative assignments then become final.

Executing DA (see Table \ref{tab:rejectionsDA}) for the problem in Table~\ref{tab:example} yields the matching:
\[
\mu^{DA} = (i_1\text{--}s_3,\; i_2\text{--}s_4,\; i_3\text{--}s_1,\; i_4\text{--}s_2),
\]
with rank profile $(2,2,4,3)$.
In this example, $\mu^{DA}$ is stable but not Pareto efficient: students $i_2$ and $i_4$ can exchange their assignments and both strictly improve without affecting any other student.\footnote{This exchange violates priorities at the relevant schools, which is why the improved matching is not stable.}

\begin{table}[h!]
	\centering
	\caption{DA proposal rounds (tentative holders shown; rejections in bold).}
	\label{tab:rejectionsDA}
	\begin{tabular}{CCCCC}
		\toprule
		& s_1 & s_2 & s_3 & s_4 \\
		\midrule
		\text{Round 1} &  & i_1,\pmb{i_2} &  & \pmb{i_3}, i_4 \\
		\text{Round 2} &  & i_1 & i_3 & i_2, \pmb{i_4} \\
		\text{Round 3} &  & i_1 & \pmb{i_3}, i_4 & i_2 \\
		\text{Round 4} &  & \pmb{i_1}, i_3 & i_4 & i_2 \\
		\text{Round 5} &  & i_3 & i_1, \pmb{i_4} & i_2 \\
		\text{Round 6} &  & \pmb{i_3}, i_4 & i_1 & i_2 \\
		\text{Round 7} & i_3 & i_4 & i_1 & i_2 \\
		\bottomrule
	\end{tabular}
\end{table}

\paragraph{Efficiency-Adjusted Deferred Acceptance (EADA).}
We explain to subjects a simplified EADA implementation following \citet{tang2014new}.
The procedure iterates as follows: run DA, identify the set of \emph{underdemanded} schools (schools that never reject an applicant during the DA execution), and \emph{permanently} match each underdemanded school with the student it holds at the DA outcome.
Remove these schools and the matched students from the problem, and repeat on the reduced instance until all students are assigned.
In the example, in the first DA run the only underdemanded school is $s_1$, which is matched to $i_3$; we remove $(i_3,s_1)$ and rerun DA on the reduced problem with students $\{i_1,i_2,i_4\}$ and schools $\{s_2,s_3,s_4\}$.

\begin{table}[h!]
	\caption{Second step in EADA: reduced preferences (left) and DA rounds (right).}
	\label{tab:seada1}
	\begin{minipage}{.52\linewidth}
		\centering
		\begin{tabular}{CCCCCCC}
			\toprule
			i_1 & i_2 & i_4 & &  s_2 & s_3 & s_4 \\
			\midrule
			s_2 & s_2 & s_4 & & i_4 & i_2 & i_1 \\
			s_3 & s_4 & s_3 & & i_1 & i_1 & i_2 \\
			s_4 & s_3 & s_2 & & i_2 & i_4 & i_4 \\
			\bottomrule
		\end{tabular}
	\end{minipage}\hfill%
	\begin{minipage}{.44\linewidth}
		\centering
		\begin{tabular}{CCCC}
			\toprule
			& s_2 & s_3 & s_4 \\
			\midrule
			\text{Round 1} & i_1,\pmb{i_2} &  & i_4 \\
			\text{Round 2} & i_1 &  & i_2,\pmb{i_4} \\
			\text{Round 3} & i_1 & i_4 & i_2 \\
			\bottomrule
		\end{tabular}
	\end{minipage}
\end{table}

In this reduced DA run, $s_3$ is underdemanded and is matched to $i_4$; we remove $(i_4,s_3)$ and rerun DA on the remaining problem with students $\{i_1,i_2\}$ and schools $\{s_2,s_4\}$.

\begin{table}[h!]
	\caption{Third step in EADA: reduced problem (left) and DA rounds (right).}
	\label{tab:seada2}
	\begin{minipage}{.52\linewidth}
		\centering
		\begin{tabular}{CCCCC}
			\toprule
			i_1 & i_2 & & s_2 & s_4 \\
			\midrule
			s_2 & s_2 & & i_1 & i_1 \\
			s_4 & s_4 & & i_2 & i_2 \\
			\bottomrule
		\end{tabular}
	\end{minipage}\hfill%
	\begin{minipage}{.44\linewidth}
		\centering
		\begin{tabular}{CCC}
			\toprule
			& s_2 & s_4 \\
			\midrule
			\text{Round 1} &i_1,\pmb{i_2} &  \\
			\text{Round 2} & i_1 & i_2 \\
			\bottomrule
		\end{tabular}
	\end{minipage}
\end{table}

In this reduced instance, $s_4$ is under-demanded and matched permanently to $i_2$. The last round trivially matches $i_1$ and $s_2$.
The resulting EADA matching is therefore:
\[
\mu^{EADA} = (i_1\text{--}s_2,\; i_2\text{--}s_4,\; i_3\text{--}s_1,\; i_4\text{--}s_3),
\]
with rank profile $(1,2,4,2)$.
In the experiment, we explain this simplified EADA implementation because it is easier to describe and to follow.\footnote{\citet{cerrone2022school} explain an implementation closer to Kesten's original formulation, where the ``interrupter(s)'' modify their preferences to avoid wasteful proposals. This implementation in general takes longer to implement, as we do not drop students after each iteration, but merely alter their preferences. It is also arguably more complex to understand. }

Part of the success behind EADA is that, as originally formulated, it first asks students ex-ante who is willing to consent to waive their schools priority when doing so does not harm them. In our experiment, we simply assume that all subjects would consent for two reasons.
First, \citet{cerrone2022school} document that different consent structures (opt-in, opt-out, or forced) yield similar conclusions in the lab (albeit with EADA being more efficient with complete consent than with observed consent, which is around 50\%).
Second, the all-consent implementation is simpler to explain to subjects.

	\paragraph{Rank-Minimizing (RM)}
	In the example, DA yields rank profile $(2,2,4,3)$ and EADA yields $(1,2,4,2)$.
	While the EADA matching is Pareto-efficient, the sum of ranks can be reduced further.
	The RM mechanism selects a matching that minimizes the sum of assigned ranks
	subject to feasibility constraints.\footnote{Operationally, RM can be computed via an assignment linear program. If multiple rank-minimizing matchings exist, RM selects one at random.} A rank-minimizing matching in our running example is
	\[
	\mu^{RM} = (i_1\text{--}s_1,\; i_2\text{--}s_2,\; i_3\text{--}s_4,\; i_4\text{--}s_3),
	\]
	which has a corresponding rank profile $(3,1,1,2)$.

	\section{Experimental Design}
	\label{sec:design}
	
	We conduct a laboratory experiment with five treatments in an incomplete factorial design.
	Three treatments correspond to the mechanisms introduced in Section~\ref{sec:model} -- DA, EADA, and RM -- with {parents} as participants.
	Parents are recruited from the the Colchester area (Essex, UK).
	Two additional treatments serve as baselines with students participating under DA, one under high stakes (the same payments we give to parents) and one under low stakes (half of the payment we give to parents).
	We use DA as the basis for comparison between students and parents because it is both the policy benchmark and the most extensively studied mechanism in the experimental school choice literature.
	The low-stakes treatment allows us to assess whether any differences between parents and students are driven by stakes rather than population characteristics, since parents' high payments substantially exceed typical experimental earnings in the school choice literature.
	
	Each experimental market consists of $18$ subjects competing for $18$ seats distributed across $7$ schools.
	Schools $s_1$ and $s_2$ each have capacity $q_{s_1}=q_{s_2}=4$, while schools $s_3$ through $s_7$ each have capacity $q_s=2$.
	
	\subsection{Preferences and Priorities}
	
	Subjects' preferences are correlated, using the ``designed'' preference environment in CS06.
	Each subject $i$ has a von Neumann-Morgenstern utility for each school $s$ given by
	\[
	u^i(s) \;=\; u^i_w(s) \;+\; u^i_q(s) \;+\; u^i_r(s),
	\]
	where:
	\begin{itemize}
		\item $u^i_w(s)=10$ if subject $i$ is in the walking zone of school $s$, and $0$ otherwise.
		Four subjects are in the walking zone of each of $s_1$ and $s_2$, and two subjects are in the walking zone of each of $s_3,\ldots,s_7$.
		
		\item Odd-index subjects satisfy $u^i_q(s_1)=40$, $u^i_q(s_2)=20$, and $u^i_q(s)=10$ for $s\in\{s_3,\ldots,s_7\}$.
		Even-index subjects satisfy $u^i_q(s_1)=20$, $u^i_q(s_2)=40$, and $u^i_q(s)=10$ for $s\in\{s_3,\ldots,s_7\}$.
		
		\item $u^i_r(s)$ is drawn uniformly on $[0,40]$ and captures heterogeneity in taste.
	\end{itemize}
	
	We draw one realization of utilities and take the induced ordinal preferences as subjects' true preferences.
	The resulting preference profile is reported in Table~\ref{tab:preferences} in Appendix \ref{app:preferencesused}. On the other hand, schools' priorities are drawn i.i.d., and one specific realization  is fixed throughout the experiment. This realization appears in Table~\ref{tab:priorities1} in Appendix \ref{app:preferencesused}.

	Under truthful reporting, the DA, EADA, and RM outcomes for this market yield average assigned ranks of 2.72, 2.44, and 1.66, respectively.
	In this environment, the DA outcome is not Pareto-efficient, while the EADA and RM outcomes are Pareto-efficient (see Appendix \ref{app:preferencesused} for the exact allocation generated by the three mechanisms).
	Moreover, relative to truthful preferences and the realized priority profile, EADA and RM generate justified envy (two and six students experience justified envy, respectively), whereas DA does not.
	Both EADA and RM can be profitably manipulated in this problem.
	The two student DA treatments differ only in the payoff scale (high vs.\ low stakes); all other features of the environment and interface are held fixed.
	
	In the parent treatments and the high-stakes student treatment, subjects earn \pounds55 if assigned to their top-ranked school and \pounds2 if assigned to their lowest-ranked school.
	Following CS06 and CHK24, we use a convex payoff schedule so that payoff differences are larger at the top of the ranking than at the bottom.
	Table~\ref{tab:payoffs} reports the payoff schedule for the high-stakes treatments.
	Subjects also receive a \pounds5 show-up fee and can earn up to \pounds4 in the Raven's matrices task that follows the main ranking task.
	
	\begin{table}[H]
		\centering
		\caption{Payoffs in the experiment (half for the low-stakes treatment).}
		\label{tab:payoffs}
		\begin{tabular}{lccccccc}
			\toprule
			School rank & 1 & 2 & 3 & 4 & 5 & 6 & 7 \\
			\midrule
			Payoff (\pounds) & 55 & 40 & 30 & 20 & 10 & 5 & 2 \\
			\bottomrule
		\end{tabular}
	\end{table}
	
	\subsection{Information and Interface}
	
\begin{figure}[H]
		\centering
		\includegraphics[width=0.8\linewidth]{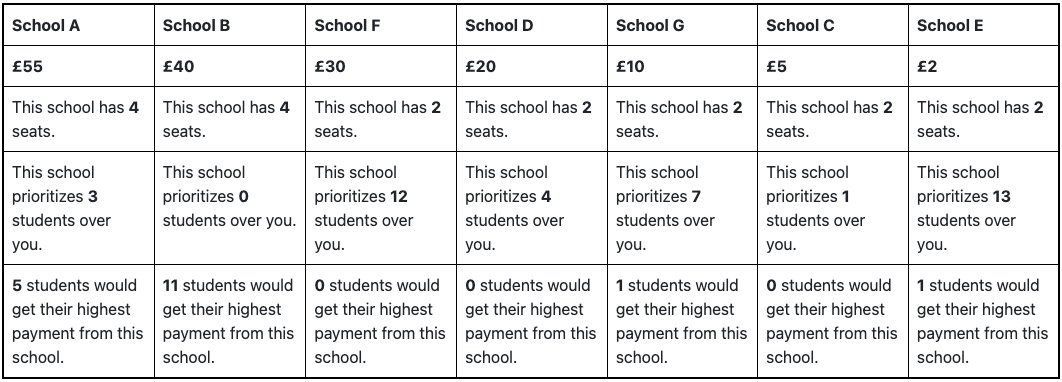}
		\caption{Information table displayed to subjects.}
		\label{fig:InterfaceInformation}
	\end{figure}

	Subjects know how much each school is worth to them (their induced preferences), their priority at each school, and can see summary information about competition, but they do not observe other participants' preferences or payoffs. For each school, an information table (Figure~\ref{fig:InterfaceInformation}) displays: (i) the monetary payoff from assignment to that school; (ii) the school's capacity; (iii) the number of other subjects who have higher priority at that school; and (iv) the number of other subjects who rank that school first.
	
	In our view, this partial information design balances realism and cognitive load. Presenting complete preference and priority tables -- as in CHK24 -- requires subjects to process two $18 \times 7$ matrices, which risks overwhelming participants and may not reflect real-world conditions where parents do not know in detail the preferences and priorities of every other parent and school. Presenting no information at all -- as in CS06 -- forces subjects to act under complete uncertainty, arguably too sparse for environments where families typically know which schools are considered popular and some notion of their acceptance chances at each school. Our design occupies a middle ground where subjects can gauge competition and their relative priority standing in a way that can be reasonably understood.
	
\begin{figure}[H]
		\centering
		\includegraphics[width=0.75\linewidth]{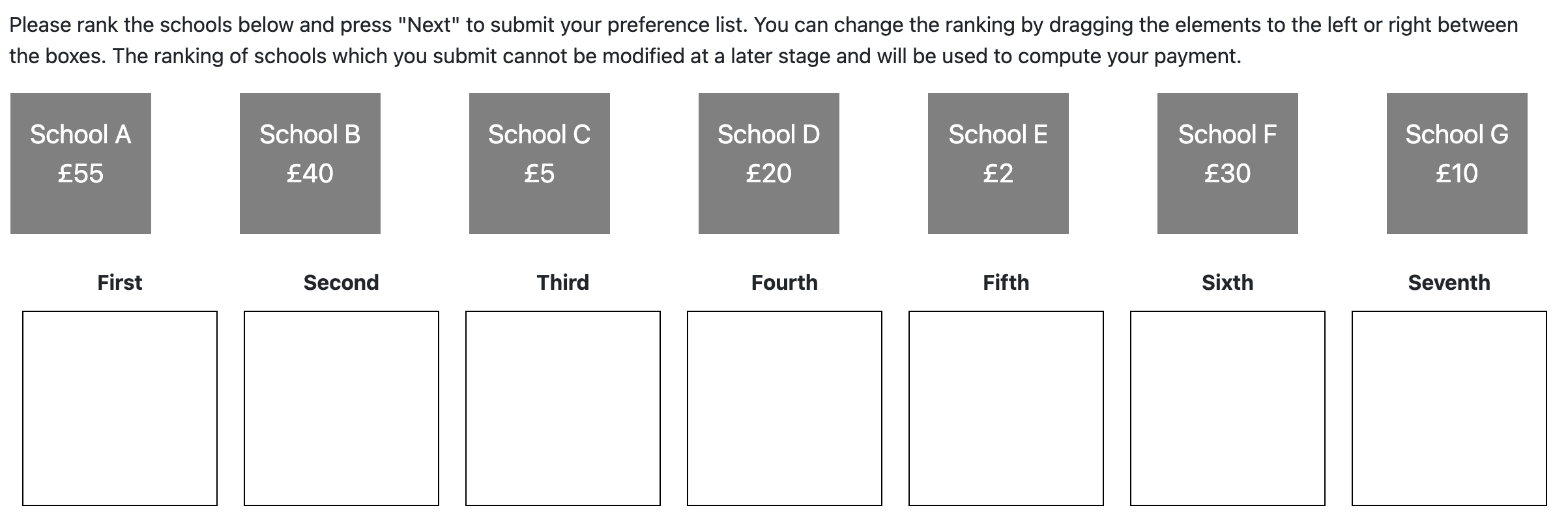}
		\caption{Ranking interface.}
		\label{fig:InterfaceDecision1}
	\end{figure}

	Figure~\ref{fig:InterfaceDecision1} shows the ranking interface. Subjects submit a complete ranking of all seven schools using a drag-and-drop tool. All subjects must rank all schools, so truncation is not allowed.
	
	Following CS06, subjects play a single round with no feedback or learning opportunities, unlike CHK24 where subjects interact repeatedly over 20 periods. Our one-shot design mirrors actual school choice: parents typically submit preferences once, without the opportunity to experiment or adjust. To ensure subjects understand the interface, we provide a non-payoff practice round in which they rank schools using the drag-and-drop tool.
	\subsection{Timeline}
	
	The experiment consists of four parts:
	\begin{enumerate}
		\item Subjects receive instructions on (i) the school choice environment and how to use the interface and (ii) the mechanism assigned to their treatment. They then complete two quiz blocks: one on the mechanism and one on the information table.\footnote{Each quiz block lasted 10 minutes. Subjects who did not complete a quiz block within the time limit proceeded to the next stage and are recorded as having failed that quiz. We also record the number of attempts required to complete each quiz.}
		\item Subjects submit their ranking of the seven schools.
		\item  Subjects complete three blocks of Raven's progressive matrices to measure cognitive ability, following \citet{basteck2018cognitive}. The Raven task lasts 18 minutes and pays up to \pounds4 depending on performance.
		\item  Subjects complete a short questionnaire collecting demographic information.
	\end{enumerate}
	
	All instructions and additional screenshots are provided in the Appendix \ref{appendix}.
	
	\subsection{Procedures}
	
	The experiment is programmed in oTree \citep{chen2016otree}.
	Sessions were conducted at 
	at ESSEXLab and Firstsite (both in Colchester, Essex).
	The experiment ran between July 2024 and July 2025.
	On average, subjects spent 33 minutes in the lab and earned \pounds38.32, inclusive of the \pounds5 show-up fee.
	
	To recruit and accommodate parents, we used an asynchronous protocol appropriate for a one-shot design.
	Participants could arrive individually during prescribed time slots, complete the study without waiting for others, and leave immediately after finishing. 
	Because outcomes require combining submissions into markets, payments could not be made on the day of participation. 
	Instead, all payments were processed digitally within two weeks. 
	To ensure credibility of delayed payments, subjects received a receipt upon completing the session guaranteeing them a minimum payment equal to the show-up fee. Subjects were informed of this procedure and the reason for the delay in advance.	

	\subsection{Sample Characteristics}
	Table~\ref{tab:demographics_summary1} reports summary statistics for our subject pools. Parents are on average 43 years old, predominantly female (about 70\%), and most have at least one child who has completed primary school. Students are younger (mean age 21-25 depending on treatment) with a more balanced gender composition.
	Our parent sample is broadly representative of the local population. Table~\ref{tab:demographics_comparison1} compares our sample to the British Household Panel Survey (BHPS 2022, restricted to East Anglia households with children) and the 2021 Census (Colchester district).
	Our sample skews slightly more female (70\% vs.\ 56\% in BHPS) but is otherwise comparable in age and marital status.
		\begin{table}[H]
		\centering
		\caption{Sample characteristics.}
		\label{tab:demographics_summary1}
		\begin{tabular}{lccccc}
			\toprule
			& \multicolumn{3}{c}{Parents} & \multicolumn{2}{c}{Students} \\
			\cmidrule(lr){2-4} \cmidrule(lr){5-6}
			& DA & EADA & RM & sDA & sDAL \\
			\midrule
			Age (mean) & 42.3 & 43.0 & 43.7 & 25.0 & 20.7 \\
			Female (\%) & 72 & 67 & 71 & 42 & 67 \\
			Children past primary (mean) & 1.15 & 1.03 & 1.18 & 0.34 & 0.21 \\
			Working hours (mean) & 28.3 & 30.3 & 27.5 & 23.4 & 15.0 \\
			\midrule
			$N$ & 108 & 108 & 108 & 108 & 108 \\
			\bottomrule
		\end{tabular}
	\end{table}

	\begin{table}[H]
	\centering
	\caption{Comparison to population benchmarks.}
	\label{tab:demographics_comparison1}
	\begin{tabular}{lccc}
		\toprule
		& Our sample & BHPS 2022 & Census 2021 \\
		& (Parents) & (E.\ Anglia w/ kids) & (Colchester) \\
		\midrule
		Female & 70\% & 56\% & 51\% \\
		Age (median) & 42 & 41 & 39 \\
		Married & 47\% & 48\% & 45\% \\
		\bottomrule
	\end{tabular}
\end{table}
	
	\subsection{Ethics}
	
	This study was approved by the University Institutional Review Board. Participation was
	voluntary, informed consent was obtained from all participants, and all analyses were conducted
	using anonymized data.
	
	\section{Hypotheses}
	\label{sec:hypotheses}

	\subsection{Truth-Telling}

	We take induced preferences as subjects' true preferences, and measure truth-telling as exact submission of the induced ranking (albeit with a number of robustness tests that will be described in turn).
	
	Since DA is strategy-proof, but EADA and RM are not, the first and most natural hypothesis is that the truth-telling rates are higher in DA than under EADA and RM.
	
	\begin{hypothesis}
		\label{hyp:truth_comparative}
		Truth-telling rates are higher under DA than under EADA or RM.
	\end{hypothesis}
	
	Hypothesis \ref{hyp:truth_comparative} reflects standard incentive reasoning but is not guaranteed empirically: the experimental literature documents widespread deviations even under strategy-proof mechanisms, and CHK24 find higher truth-telling under EADA than DA. We make no prediction about EADA versus RM.
	
	\subsection{External Validity}
	
	The experimental school choice literature relies almost exclusively on undergraduate subjects. If parents -- the natural decision-makers in school choice -- behave differently, laboratory evidence based on students may be a poor guide to policy. 
	
	Existing research on subject pool differences offers conflicting predictions. Parents are typically more risk-averse than students \citep{dohmen2011individual}, which should increase truth-telling under DA \citep{klijn2013preference}. But parents may also be more skeptical of official advice or more inclined toward strategic experimentation given their experience with high-stakes decisions. We therefore treat external validity as an empirical question rather than a directional prediction.
	
	\begin{hypothesis}
		\label{hyp:external_rates}
		Parents and students exhibit similar truth-telling rates under DA.
	\end{hypothesis}
	More broadly, even if aggregate truth-telling rates coincide, parents and students may deviate from their induced preferences in systematically different ways. We therefore also examine whether the two subject pools differ in how they deviate.
	
	\begin{hypothesis}
		\label{hyp:external_patterns}
		Parents and students exhibit similar behavioral patterns under DA.
	\end{hypothesis}

	\subsection{Mechanism Performance under Observed Behavior}
	
	Theory has clear predictions on the efficiency and stability of the allocations generated by DA, EADA and RM when subjects report their preferences truthfully, but these results may not extend to realistic cases when agents report their preferences strategically. Indeed, \citet{dur2020you} show that EADA does not Pareto-dominate DA in equilibrium. Thus, we are interested in whether the predictions generated under truthful reports hold under observed behavior in the lab. Below, we state hypotheses regarding the i) rank- and Pareto-efficiency, ii) tail outcomes, and iii) stability of the allocations generated by DA, EADA and RM. We stress that all predictions are derived under truthful behavior.
	
	\paragraph{Efficiency.}
	The leading efficiency notion in matching is Pareto-efficiency, which merely requires that no group of students is willing to exchange their placements without harming anyone. EADA and RM achieve this desideratum, which DA fails. This well-known observation gives us a natural hypothesis.
	
	\begin{hypothesis}
		\label{hyp:pareto} EADA and RM produce more Pareto-efficient allocations than DA.
	\end{hypothesis}
	
	A stronger efficiency notion is that of rank-efficiency, which means that the average rank attained by a student is as good (i.e. small) as possible. Since EADA Pareto-dominates DA, naturally we should expect EADA to generate smaller average ranks than DA, but larger than RM, which by construction represents the best rank-efficient outcome.
	
	\begin{hypothesis}
		\label{hyp:rankeff}
		Average rank is lowest in RM, followed by EADA, and highest in DA.
	\end{hypothesis}
	
	\paragraph{Stability.}
	Finally, we study the stability costs of improving rank outcomes.
	Measured relative to induced preferences and the realized priority profile, DA should exhibit the fewest priority violations, while RM should exhibit the most, with EADA sitting in the middle.\footnote{The lack of justified envy in DA is well-known. EADA is setwise envy-minimal in the class of efficient mechanisms \citep{dougan2021minimally,tang2021weak}. While this result does not mean that EADA always produces fewer blocking pairs than RM (which can happen in particular instances, see \cite{knipe2025improvable}), we expect that on average EADA produces fewer blocking pairs simply because it actually takes priorities into account.}
	
	\begin{hypothesis}
		\label{hyp:envy}
		The number of students who experience justified envy is lowest in DA, followed by EADA, and highest in RM.\end{hypothesis}
	
	\subsection{Cognitive Ability}
	
	We measure cognitive ability using Raven's Progressive Matrices, a standard nonverbal test of fluid intelligence widely used in experimental economics (e.g. \citet{basteck2018cognitive}; see \citet{sofianos2025intelligence} for an overview). Subjects complete three incentivized and timed blocks of pattern-recognition problems; we use the total score as our ability measure and classify subjects as high or low ability based on the within-pool median split.
	
	If strategic sophistication drives behavior, higher-ability subjects should either (i) recognize that truth-telling is optimal under DA, or (ii) identify profitable deviations under EADA and RM.
	
	\begin{hypothesis}
		\label{hyp:cognitive}
		Higher cognitive ability is associated with (a) higher truth-telling rates under DA and (b) lower truth-telling rates and better outcomes under EADA and RM.
	\end{hypothesis}
	
	This hypothesis has equity implications. If cognitive ability predicts outcomes under EADA and RM but not under DA, then strategy-proof mechanisms may be valuable not because they induce truth-telling, but because they prevent sophisticated families from gaining advantage through strategic play.
	
	\section{Results}
	\label{sec:results}

	\subsection{Truth-telling and External Validity}
	
	Table~\ref{tab:truthtelling_main} presents truth-telling rates across all treatments.
	The first row reports the share of subjects who submitted preferences identical to their induced (true) preferences.
	Contrary to Hypothesis~\ref{hyp:truth_comparative}, we find substantial misreporting in all mechanisms, including the strategy-proof DA.
	Only 26\% of parents report truthfully under DA, with similar rates under EADA (29\%) and RM (23\%).
	Students also display high manipulation rates: 26\% report truthfully under high-stakes DA (sDA) and 18\% under low-stakes DA (sDAL).
	
	\begin{table}[H]
		\centering
		\caption{Truth-telling rates.}
		\label{tab:truthtelling_main}
		\begin{tabular}{lccccc}
			\toprule
			& \multicolumn{3}{c}{Parents} & \multicolumn{2}{c}{Students} \\
			\cmidrule(lr){2-4} \cmidrule(lr){5-6}
			& DA & EADA & RM & sDA & sDAL \\
			\midrule
			Truth-telling rate & 0.26 & 0.29 & 0.23 & 0.26 & 0.18 \\
			\bottomrule
			\multicolumn{6}{l}{\footnotesize $N = 108$ per treatment (6 independent markets $\times$ 18 subjects).}
		\end{tabular}
	\end{table}
	
	\begin{result}
		\label{res:truth}
		Hypothesis~\ref{hyp:truth_comparative} is rejected: truth-telling is not higher under DA than under EADA or RM  (Fisher’s exact tests, $p=0.76$ for DA vs.\ EADA and $p=0.75$
		for DA vs.\ RM).\end{result}
	
	Pairwise Fisher’s exact tests also fail to detect any differences in truth-telling rates between
	EADA and RM ($p=0.44$).
	
	The truth-telling rates are relatively low compared to CS06 and CHK24 (but see also  \citet{guillen2021strategy} for much lower truthful levels). In the following subsections, we will show that  Result \ref{res:truth} is not caused by a poor understanding of the mechanisms, and no longer holds when we use a weaker definition of truthfulness: we find significantly fewer consequential manipulations in strategy-proof DA.

	Regarding external validity, we find that truth-telling rates are nearly identical for parents and students (we use the comparison between DA versus pooled sDA and sDAL, reporting test for pairwise comparisons below).
	Under low stakes, students truth-tell somewhat less (18\%), though the magnitude is not large.
	
	\begin{result}
		\label{res:extvalid}
		Consistent with Hypothesis~\ref{hyp:external_rates}, parents and students exhibit similar aggregate manipulation rates under DA (Fisher's exact test $p = 0.41$).
	\end{result}
	
	The difference between truth-telling rates in DA versus sDA or versus sDAL is not statistically significant in either comparison (Fisher's exact test against sDA: p = 1.00; against sDAL: p = 0.19). The absence of significant stakes effects among students suggests that the similarity between parents and students in aggregate manipulation rates is not an artifact of the high-stakes design.
	Result \ref{res:extvalid} supports the external validity of the large experimental school choice literature that relies on student subjects, at least for aggregate DA manipulation rates in a CS06-style correlated-preference environment.
	
	\paragraph{Do subjects understand the mechanism?}
	
	A natural explanation for the observed low truth-telling across all treatments is that subjects did not understand the mechanisms.
	We present evidence against this interpretation.
	
	Before making their ranking decisions, subjects completed a comprehension quiz on the mechanism.
	Table~\ref{tab:quiz} reports quiz performance across treatments.
	Approximately 88\% of subjects successfully completed the quiz within the 10-minute time limit, with no large differences across mechanisms.
	Furthermore, about two-thirds of subjects passed on their first attempt.
	
	\begin{table}[H]
		\centering
		\caption{Quiz performance by mechanism}
		\label{tab:quiz}
		\begin{tabular}{lccccc}
			\toprule
			& \multicolumn{3}{c}{Parents} & \multicolumn{2}{c}{Students} \\
			\cmidrule(lr){2-4} \cmidrule(lr){5-6}
			& DA & EADA & RM & sDA & sDAL \\
			\midrule
			Passed quiz (within 10 min) & 0.88 & 0.87 & 0.89 & 0.94 & 0.98 \\
			Passed on first attempt & 0.66 & 0.68 & 0.65 & 0.61 & 0.82 \\
			Mean attempts & 2.76 & 2.26 & 2.56 & 3.39 & 1.79 \\
			\bottomrule
		\end{tabular}
	\end{table}
	
	The mean number of quiz attempts (2.26 to 3.39) reflects a skewed distribution: many subjects pass quickly, while a small number require many attempts.
	Importantly, subjects could proceed only after passing, so quiz completion indicates a basic grasp of the mechanism as presented.

	\paragraph{Obvious mistakes.}
	A more direct test of basic strategic understanding is whether subjects avoid dominated choices when the environment makes the optimal action transparent.
	We define an \emph{obvious mistake} as failing to rank one's true first choice first when that school is a \emph{safe school}, i.e. one where the subject's priority rank is no worse than the school's capacity (i.e., priority rank $\le q_s$).
	For four subject positions in our design ($i_1$, $i_6$, $i_{12}$, and $i_{15}$), the induced first choice is safe.
	For these subjects, ranking any other school first is a clear error under both DA and EADA (but not under RM, where having a high priority is irrelevant for one's assignment).
	
	Table~\ref{tab:obvious} reports the frequency of obvious mistakes.
	Parents rarely make this error (4\% in DA and 13\% in EADA), while students make it much more frequently (42\% in sDA and 25\% in sDAL).
	This difference is large and statistically meaningful in magnitude.
	
	\begin{result}
		\label{res:obviousmistakes}
		Students make obvious mistakes significantly more often than parents 
		(33\% vs 8\%; Fisher's exact test pooled $p = 0.005$).
	\end{result}
	
	Result \ref{res:obviousmistakes} also obtains when comparing only obvious mistakes in DA ( 4\% of parents make this error, compared to 33\% of 
	students pooled across stakes conditions, Fisher's exact test 
	$p = 0.007$).
	Result~\ref{res:obviousmistakes} implies that, while aggregate truth-telling rates are similar, the nature of deviations differs systematically across subject pools.
	
	\begin{result}
		\label{res:extvalid_patterns}
		Hypothesis~\ref{hyp:external_patterns} is rejected: parents and students exhibit different behavioral patterns under DA. Students' deviations are more consistent with avoidable mistakes, whereas parents' deviations appear more deliberate.
	\end{result}
	
	\begin{table}[H]
		\centering
		\caption{Obvious mistakes by mechanism (among subjects with safe first choice)}
		\label{tab:obvious}
		\begin{tabular}{lcccc}
			\toprule
			& \multicolumn{2}{c}{Parents} & \multicolumn{2}{c}{Students} \\
			\cmidrule(lr){2-3} \cmidrule(lr){4-5}
			& DA & EADA & sDA & sDAL \\
			\midrule
			Absolute & 1 & 3 & 10 & 6 \\
			Relative & 0.04 & 0.13 & 0.42 & 0.25 \\
			\midrule
			$N$ & 24 & 24 & 24 & 24 \\
			\bottomrule
		\end{tabular}
	\end{table}

	\paragraph{Consequential manipulations.}
	
	Obvious mistakes lead to worse payoffs, and thus are one type of consequential manipulations, i.e. non-truthful reports that change the manipulator's payoff. As a robustness test for Results \ref{res:truth} and \ref{res:extvalid},  we  measure how many {consequential manipulations} are observed in each of the mechanisms (consequential holding all other subjects' submitted reports fixed).
	Table~\ref{tab:consequential} reports the share of subjects whose deviations are consequential under this definition (among all subjects, not just manipulators).\footnote{RM produces a rank-efficient allocation, and since it need not be unique, it may choose one randomly. Thus, whether the report is consequential or profitable may be due to luck only. In parenthesis, we report the corresponding statistic computed over all potential rank efficient outcomes for robustness. Using this alternative measure does not change the results significantly.} Students display slightly higher consequential manipulation rates than parents under DA, but the difference is not statistically significant.
	
	\begin{table}[H]
		\centering
		\caption{Consequential manipulations.}
		\label{tab:consequential}
		\begin{tabular}{cccccc}
			\toprule
			&\multicolumn{3}{c}{Parents} & \multicolumn{2}{c}{Students} \\
			&DA & EADA & RM  & sDA & sDAL \\
			\midrule
		  Absolute &14 &  29 &  51 (68)$^*$ &  20 &  23\\ 
		Relative&	0.13 & 0.27 & 0.47 (0.63)$^*$ & 0.19 & 0.21 \\
	\midrule
$N$ & 108 & 108 & 108 & 108&108 \\
			\bottomrule
			\multicolumn{6}{l}{$^*$Average across all rank-minimizing allocations.}
		\end{tabular}
	\end{table}
	
	A clear pattern emerges: consequential manipulation is far more common under RM (47\%) than under DA (13\%) or EADA (27\%).
	This matches the spirit of Hypothesis \ref{hyp:truth_comparative}, showing that while manipulations are similar across all three treatments, strategy-proof DA gets significantly fewer consequential manipulations than EADA and RM.
	
	\begin{result}
		\label{res:conseq}
		Consequential manipulations are significantly more prevalent under RM than under
		EADA, and more prevalent under EADA than under DA (Fisher’s exact tests: RM vs.\ DA,
		$p<0.001$; RM vs.\ EADA, $p=0.003$; EADA vs.\ DA, $p=0.016$).  Compared to parents in DA,
		students in DA exhibit a higher rate of consequential manipulations, but these differences are not statistically significant (DA vs.\ pooled sDA+sDAL, $p=0.163$).
	\end{result}
	

	We have seen that DA generates significantly fewer consequential manipulations than EADA and RM. But are these fewer manipulations ever profitable?
	Under DA, strategy-proofness implies that any deviation from truthful reporting cannot yield a strictly better assignment for the deviator, holding others' reports fixed.
	Under EADA and RM, by contrast, some consequential deviations may be beneficial. We therefore decompose consequential manipulation into:
	(i) consequential-and-beneficial and
	(ii) consequential-and-harmful. 
	
	Interestingly, we find that no manipulator ever benefits in DA, nor in EADA. In fact, 13\% and 27\% of subjects manipulate their reported preferences and end with a worse outcome because of this. The corresponding fraction for RM is much higher: 41\%, yet this is the only mechanism where some manipulations pay off: 6\% of subjects manipulate their preferences successfully, as we report in Table \ref{tab:consequential2} below. Importantly, we emphasize that opportunities for successful manipulations in EADA are available, yet no subject successfully manipulates EADA. We summarize these observations in Result \ref{res:negativemanipulation}.\footnote{One such manipulation is for subject $i_7$ to report the preference profile $s_1\succ s_2 \succ s_5 \succ s_3 \succ s_7 \succ s_6 \succ s_4$, which moves him from his third to his second-best school.} 
	
	\begin{result}
		\label{res:negativemanipulation}
		Harmful  manipulations are significantly more prevalent under RM than
		under EADA, and more prevalent under EADA than under DA (Fisher’s exact tests: RM vs.\
		DA, $p<0.001$; RM vs.\ EADA, $p=0.044$; EADA vs.\ DA, $p=0.016$). Yet, only in RM 
		are some manipulations successful at a rate significantly greater than zero 
		(binomial test $p<0.001$).
	\end{result}

	\begin{table}[H]
		\centering
		\caption{Decomposition of consequential manipulations.}
		\label{tab:consequential2}
		\begin{tabular}{lccccc}
			\toprule
			& \multicolumn{3}{c}{Parents} & \multicolumn{2}{c}{Students} \\
			\cmidrule(lr){2-4} \cmidrule(lr){5-6}
			& DA & EADA & RM & sDA & sDAL \\
			\midrule
			Share with beneficial manipulation & 0 & 0 & 0.06 (0.10)$^*$ & 0 & 0 \\
			Average rank gain & 0 & 0 & 1.14  (0.53)$^*$ & 0 & 0 \\
			\hline
			Share with harmful manipulation & 0.13 & 0.27 & 0.41  (0.53)$^*$ & 0.19 & 0.21 \\
			Average rank loss & 2.36 & 1.76 & 2.43 (1.56)$^*$ & 1.95 & 1.74 \\
			\bottomrule
			\multicolumn{6}{l}{$^*$ Average across all rank-minimal allocations.}
		\end{tabular}
	\end{table}
	
	\paragraph{Further analysis.}
	We report additional analyses in Appendix~\ref{app:further} on the types 
	of deviations subjects employ. Two recurring patterns emerge across all 
	mechanisms: skipping down (demoting schools perceived as unattainable) 
	and inflating demand (promoting a popular school into a high position). 
	Both patterns occur at similar rates across mechanisms, though skipping 
	down is substantially more common among students than parents, providing 
	further support for Result~\ref{res:extvalid_patterns}. We also find 
	that the position at which misreporting first occurs is similar across 
	treatments; formal tests fail to reject equality of the first-deviation 
	distributions ($\chi^2$ test of independence, $p=0.88$).
	
	In summary, misreporting is similarly common under DA, EADA, and RM, 
	and subjects employ similar heuristics regardless of the mechanism. 
	What differs across mechanisms is the impact of deviations: in the 
	non-strategy-proof mechanisms, deviations are more likely to be 
	consequential. What differs across subject pools is the nature of 
	deviations: parents and students misreport at similar rates 
	(Result~\ref{res:extvalid}), but students commit substantially more 
	obvious errors and rely more heavily on naive heuristics 
	(Result~\ref{res:extvalid_patterns}).
	
\subsection{Mechanism Performance}

We now turn to mechanism performance under observed behavior. Despite widespread manipulation
under all mechanisms, do the predicted efficiency-stability tradeoffs survive? We examine
Pareto-efficiency, rank-efficiency and stability
 (H\ref{hyp:pareto}-\ref{hyp:envy}) using the recombinant estimation procedure described below.

Our unit of independence is the market (an 18-subject group), and we observe six independent
markets per treatment. To characterize mechanism performance at the market level and obtain
uncertainty measures, we use the recombinant estimation approach \citep{mullin2006recombinant},
as in CS06. Each recombination draw constructs a synthetic market by sampling, for each of the
18 subject positions, that position's submitted ranking from one of the six observed markets
within the same treatment. We then run the mechanism on the synthetic market and compute the
corresponding market outcomes. Repeating this procedure 10{,}000 times per treatment yields an
estimated distribution of each performance statistic and associated confidence intervals. We
also report a robustness check based on 200{,}000 recombinations. Details appear in
Appendix~\ref{app:recomb}.

\paragraph{Pareto-efficiency.}
\label{sec:paretofreq}
Under truthful reporting, EADA and RM guarantee Pareto-efficient allocations, while DA need not. Under observed behavior, however, no mechanism is guaranteed to be Pareto-efficient relative to induced preferences. We therefore take as our primary outcome the frequency with which each mechanism yields a Pareto-efficient allocation in the induced preference profile.

\begin{figure}[H]
	\centering
	\caption{Frequency of Pareto-efficient allocations by mechanism (recombinant estimates with 95\% CI)}
	\label{fig:paretofreq}
	\includegraphics[width=0.4\textwidth]{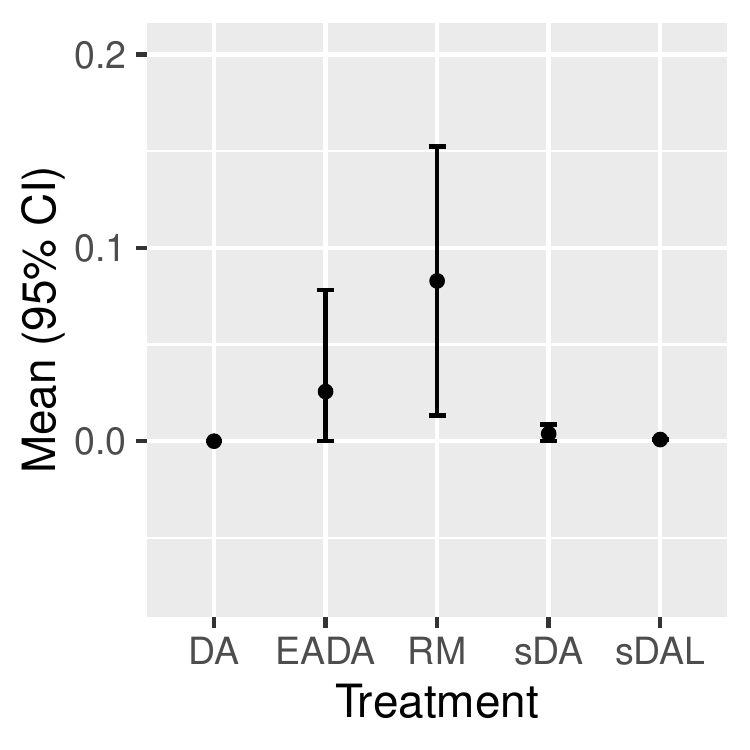}
\end{figure}

Figure~\ref{fig:paretofreq} presents recombinant estimates of the probability that the realized allocation is Pareto-efficient. RM achieves Pareto-efficiency in 8.3\% of markets, compared to 2.6\% for EADA and effectively 0\% for DA and the student treatments. Only RM delivers Pareto-efficient allocations with a frequency that is statistically distinguishable from zero. RM significantly outperforms DA (recombinant $p = 0.010$), while the difference between RM and EADA is not statistically significant ($p = 0.106$). EADA's gains over DA are modest and not statistically distinguishable from zero ($p = 0.170$).\footnote{Inference is based on the empirical distribution of recombinant draw-by-draw differences, which yields a permutation-style test that does not rely on independence or normality assumptions; classical $t$-tests are therefore inappropriate in this setting.}

\begin{result}
	\label{res:pareto}
	Consistent with Hypothesis~\ref{hyp:pareto}, RM increases the frequency of Pareto-efficient outcomes relative to DA (recombinant $p = 0.010$). EADA yields modest gains over DA that remain statistically insignificant ($p = 0.170$).
\end{result}

A natural question is why the rate at which EADA's outcome is Pareto-efficient in our experiment differs so sharply from those reported in CHK24, who document Pareto-efficient assignments in 80\% of markets when all students consent. While many design differences exist (one-shot vs.\ repeated interaction, partial vs. full information, 5 versus 18 subjects, etcetera), we argue that differences in truth-telling rates can account for most of the gap. Using the same recombinant procedure, we construct synthetic markets in which we randomly replace a fraction of submitted rankings with the induced true preferences to artificially increase the fraction of subjects who behave truthfully closer to those in CHK24 (holding all other features fixed) and re-estimate the fraction of markets that achieve Pareto-efficiency under EADA. These appear in Table \ref{tab:synthetic-pareto}.

\begin{table}[H]
	\centering
	\caption{Truth-telling and Pareto-efficiency under EADA (synthetic markets)}
	\label{tab:synthetic-pareto}
	\begin{tabular}{lc}
		\toprule
		Truth-telling rate          & Share of Pareto-efficient allocations \\
		\midrule
		0.29 (observed)             & 0.026 \\
		0.30             & 0.028 \\
		0.50           & 0.118 \\
		0.70             & 0.332 \\
		0.90             & 0.722 \\
		\bottomrule
	\end{tabular}
\end{table}
The relationship between truth-telling and Pareto-efficiency is monotone. Raising truth-telling from our observed 29\% to the 70\% level increases the probability of a Pareto-efficient allocation significantly, so that one in three markets reach efficiency. When truth-telling increases to 90\%, efficiency becomes the modal outcome (72.2\%). Thus, while other factors may also play a role, the low truth-telling rate in our parent sample appears to be a primary driver of EADA's muted performance relative to prior work.\footnote{In particular, learning from repeated play in CHK24 (versus our one-shot design) is unlikely to explain the gap: truth-telling rates under all EADA variants in CHK24 remain flat across their 20 periods.} Whether most parents would behave as in CHK24 or as in our experiment (or even exhibit further differences) is an interesting question for future work.\footnote{Synthetic markets replace a random fraction of submitted rankings with true preferences before recombination (10,000 draws per cell).}

Going back to our experiment, we also examine the fraction of improvable subjects, i.e., those who could improve their placement by exchanging schools in a Pareto trade without harming any other subject. This measure captures the scope for efficiency adjustment: higher values indicate more Pareto-improving trades remain available from the realized allocation. Point estimates suggest fewer potential Pareto improvements under RM (0.28) than under EADA (0.30) or DA (0.35). However, pairwise recombinant comparisons indicate that these differences are not statistically distinguishable from zero at conventional levels (all pairwise $p$-values $> 0.10$).
\paragraph{Rank-efficiency.}

\begin{figure}[H]
	\centering
	\caption{Average assigned rank by mechanism (recombinant estimates with 95\% CI)}
	\label{fig:avgrank}
	\includegraphics[width=0.4\textwidth]{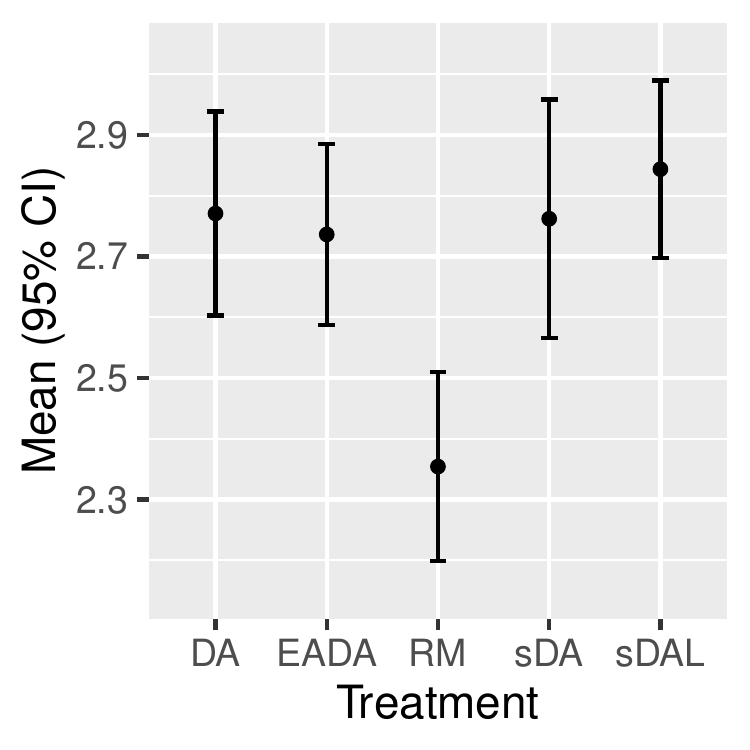}
\end{figure}

Figure~\ref{fig:avgrank} presents the average assigned rank across mechanisms, computed using recombinant estimation (lower is better). RM achieves a substantially lower average rank (2.35) than DA (2.84) and EADA (2.74).
This difference is economically meaningful in a seven-school environment.

\begin{result}
	\label{res:rankeff}
	Hypothesis~\ref{hyp:rankeff} is supported: RM delivers significantly better ranks than both DA (recombinant $p < 0.001$) and EADA (recombinant $p = 0.001$). EADA offers no significant improvement over DA (recombinant $p = 0.37$). 
\end{result}

\paragraph{Maximum Rank.} Beyond the average rank, we also analyze the maximum rank (i.e. the placement received by the worst-off student in each market).
RM delivers the best tail outcomes, with an expected maximum rank of 5.86 compared to 6.80 for DA and 6.94 for EADA.
This is consistent with RM achieving an average maximum rank lower than DA \citep{ortega2023cost}. We remark that, under truthfulness,  EADA never generates a larger maximum rank than DA (since it Pareto dominates it). In this case, since maximum ranks are obtained by different subjects with different behavior, EADA actually generates a worse worst-off placement than DA, which we find worth emphasizing. We summarize these observations in Result \ref{res:maxrank} below.
\begin{result}
	\label{res:maxrank}
	RM reduces the maximum rank by 0.94 relative to DA (recombinant $p < 0.001$) and by 1.08 relative to EADA (recombinant $p < 0.001$). EADA's maximum rank is not statistically different from DA's (mean difference $0.14$, recombinant $p = 0.15$).
\end{result}

\paragraph{Justified envy.}

DA's main appeal is stability with respect to reported preferences: given any submitted profile, DA returns a matching that is stable relative to that profile.
EADA and RM may generate priority violations, and among Pareto-efficient mechanisms, EADA minimizes  justified envy in a setwise manner \citep{dougan2021minimally,tang2021weak}.

Table~\ref{tab:envy_both} reports justified envy (i.e. blocking pairs) with respect to reported and true preferences.
Under reported envy, DA yields zero by construction, while EADA and RM generate positive rates.
Under true envy, DA need not yield zero because misreports can place students into assignments they truly dislike, while lower-priority students receive schools they truly prefer.

\begin{table}[H]
	\centering
	\caption{Justified envy under submitted and true preferences (recombinant estimates)}
	\label{tab:envy_both}
	\begin{tabular}{lccccc}
		\toprule
		& DA & EADA & RM & sDA & sDAL \\
		\midrule
		Reported& 0.00 & 0.10 & 0.29 & 0.00 & 0.00 \\
		True & 0.14 & 0.29 & 0.51 & 0.20 & 0.20 \\
		\bottomrule
	\end{tabular}
\end{table}

\begin{result}
	\label{res:envy}
	Consistent with Hypothesis~\ref{hyp:envy},  justified envy is lowest under DA (14\%), followed by EADA (29\%) and RM (51\%). EADA  generates significantly more justified envy than DA (recombinant $p < 0.001$). RM generates significantly more justified envy than EADA (recombinant $p < 0.001$). 
\end{result}

\begin{figure}[H]
	\centering
	\caption{Share of students with true justified envy by mechanism (recombinant estimates with 95\% CI)}
	\label{fig:justenvy}
	\includegraphics[width=0.4\textwidth]{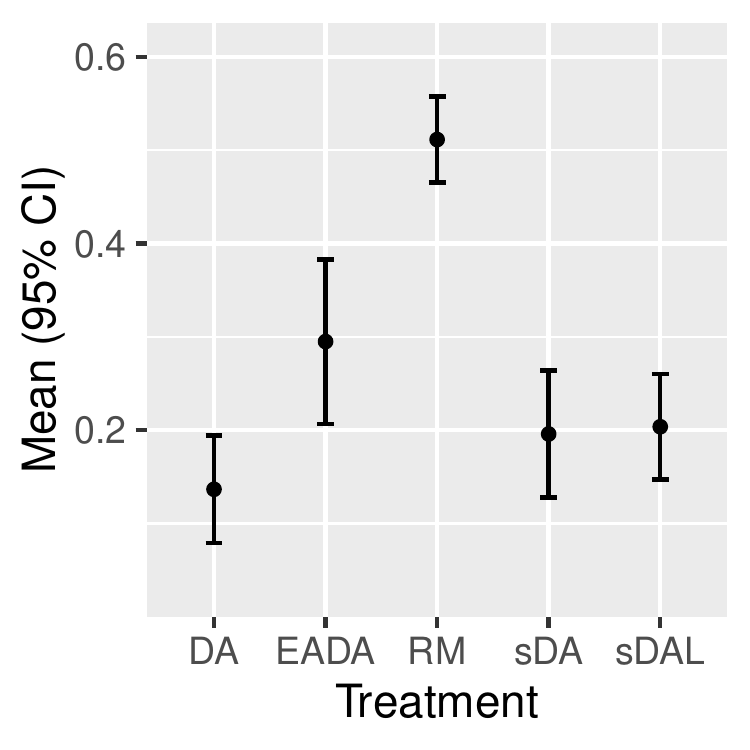}
\end{figure}

\paragraph{Summary.}

Table~\ref{tab:mechanism_summary} summarizes performance across all dimensions. 
The results reveal a stark efficiency-stability tradeoff, but an asymmetric one across mechanisms. 
DA generates significantly less justified envy than unstable mechanisms but never produces Pareto-efficient allocations and generates significantly worse rank distributions.
RM delivers substantial efficiency gains: average rank improves by nearly half a position relative to DA, and worst-off outcomes improve by almost a full rank.
These gains come at a cost: half of students experience justified envy.
EADA, by contrast, offers small and imprecisely estimated improvements relative to DA, yet it still generates twice as much justified envy than DA.
In this sense, EADA occupies a middle ground: it sacrifices DA's stability guarantee by creating twice as many blocking pairs, but does not deliver significant rank-efficiency gains nor it generates Pareto-efficient allocations at a significantly higher rate.

\begin{table}[H]
	\centering
	\caption{Mechanism performance summary (recombinant estimates)}
	\label{tab:mechanism_summary}
	\begin{tabular}{lccccc}
		\toprule
		& DA & EADA & RM & sDA & sDAL \\
		\midrule
		Pareto-efficient  & 0.00 & 0.03 & 0.09 & 0.00 & 0.00 \\
		Average rank & 2.84 & 2.74 & 2.35 & 2.78 & 2.82 \\
		Maximum rank & 6.80 & 6.94 & 5.86 & 6.55 & 6.75 \\
		
		Justified envy & 0.14 & 0.29 & 0.51 & 0.20 & 0.20 \\
		
		\bottomrule
	\end{tabular}
	\begin{minipage}{0.95\textwidth}
		\footnotesize
		\centering\textit{Note:} Based on 10,000 recombinant draws per treatment. \end{minipage}
\end{table}

\subsection{Cognitive Ability}
\label{subsec:ability}

Hypothesis~\ref{hyp:cognitive} states that higher-ability subjects should (i) be more likely to
recognize that truth-telling is a weakly dominant strategy in DA and therefore report truthfully at
higher rates, and (ii) better recognize and exploit the scope for manipulation under EADA and RM,
leading to better outcomes. We proxy cognitive ability using subjects' scores on a Raven progressive
matrices task.\footnote{Raven matrices come in blocks of 12 questions, each block progressively more difficult. We ask subjects to complete blocks C, D and E. Due to a programming error, in treatment sDA only block C answers were recorded. Our main analysis therefore uses block C scores, which allows us to retain all five treatments ($N=540$). In Appendix~\ref{app:cde}, we present an alternative specification using the full CDE score that drops treatment sDA ($N=432$). All results are directionally consistent across specifications; we flag below where statistical significance differs.}

Table~\ref{tab:cognitive1}  reports results from two specifications for each outcome: a pooled model and
a model with mechanism$\times$Raven interactions. All models use DA as the reference category. Columns (1)-(2) estimate logit models of truth-telling; columns (3)-(4) estimate OLS regressions of own assigned rank (lower is better), with standard errors clustered at the market level.

\begin{equation}
	\label{eq:logit_truth}
	\text{Truth}_{im}
	= \alpha + \beta\,\text{Raven}_{im}
	+ \sum_{k \neq \text{DA}} \gamma_k\,\mathbbm{1}\{M_{im}=k\}
	+ \sum_{k \neq \text{DA}} \delta_k\,\text{Raven}_{im} \times \mathbbm{1}\{M_{im}=k\}
	+ \varepsilon_{im},
\end{equation}

\begin{eqnarray}
	\label{eq:ols_rank}
	\nonumber \text{Rank}_{im}
	=& a + b\,\text{Raven}_{im}
	+ \sum_{k \neq \text{DA}} c_k\,\mathbbm{1}\{M_{im}=k\}
	+ d\,\overline{\text{Raven}}_{-i,m}\\
	+& \sum_{k \neq \text{DA}} e_k\,\text{Raven}_{im} \times \mathbbm{1}\{M_{im}=k\}
	+ u_{im}.
\end{eqnarray}

In the pooled specification (column 1), higher Raven scores are associated with a significantly
lower probability of submitting the induced ranking ($\hat\beta=-0.105$, $p=0.005$), contrary to
Hypothesis~\ref{hyp:cognitive}(i). Among the mechanism indicators, only sDAL differs significantly
from DA.
\begin{table}[H]
	\centering
	\caption{Cognitive ability, truth-telling, and outcomes}
	\label{tab:cognitive1}
	\small
 \begin{tabular}{l cc cc}
	\toprule
	& \multicolumn{2}{c}{Truth-telling (logit)} & \multicolumn{2}{c}{Assigned rank (OLS)} \\
	\cmidrule(lr){2-3} \cmidrule(lr){4-5}
	& (1) & (2) & (3) & (4) \\
	\midrule
	Raven & $-0.105^{***}$ & $-0.164^{*}$ & $-0.021$ & $-0.073^{**}$ \\
	& $(0.038)$ & $(0.084)$ & $(0.023)$ & $(0.033)$ \\[4pt]
	EADA & $0.106$ & $0.757$ & $-0.086$ & $-1.332^{***}$ \\
	& $(0.309)$ & $(0.985)$ & $(0.156)$ & $(0.496)$ \\[4pt]
	RM & $-0.118$ & $-2.797^{**}$ & $-0.347^{**}$ & $0.043$ \\
	& $(0.319)$ & $(1.287)$ & $(0.175)$ & $(0.508)$ \\[4pt]
	sDA & $0.045$ & $-1.099$ & $-0.092$ & $-0.306$ \\
	& $(0.313)$ & $(1.221)$ & $(0.126)$ & $(0.690)$ \\[4pt]
	sDAL & $-0.435$ & $-1.390$ & $0.138$ & $-0.548$ \\
	& $(0.338)$ & $(1.199)$ & $(0.158)$ & $(0.674)$ \\[4pt]
	Raven $\times$ EADA & & $-0.083$ & & $0.144^{***}$ \\
	& & $(0.112)$ & & $(0.054)$ \\[4pt]
	Raven $\times$ RM & & $0.296^{**}$ & & $-0.041$ \\
	& & $(0.136)$ & & $(0.049)$ \\[4pt]
	Raven $\times$ sDA & & $0.130$ & & $0.026$ \\
	& & $(0.133)$ & & $(0.068)$ \\[4pt]
	Raven $\times$ sDAL & & $0.109$ & & $0.076$ \\
	& & $(0.129)$ & & $(0.063)$ \\[4pt]
	$\overline{\text{Raven}}_{-i}$ & & & $-0.053$ & $-0.055$ \\
	& & & $(0.080)$ & $(0.067)$ \\[4pt]
	Constant & $-0.139$ & $0.367$ & $3.474^{***}$ & $3.959^{***}$ \\
	& $(0.390)$ & $(0.746)$ & $(0.748)$ & $(0.730)$ \\
	\midrule
	$N$ & 540 & 540 & 540 & 540 \\
	\bottomrule
\end{tabular}
	\vspace{4pt}
	
	{\footnotesize Notes: Raven denotes the block C score (see text). Columns (1)-(2) report logit coefficients with standard errors in parentheses; columns (3)-(4) report OLS estimates with standard errors clustered at the market level in parentheses. $^{*}p<0.10$, $^{**}p<0.05$, $^{***}p<0.01$. DA is the reference category. $\overline{\text{Raven}}_{-i}$ is included only in the rank regressions, as it should not affect one's truth-telling decision.}
\end{table}

The interaction model (column 2) yields the same qualitative conclusion. Under DA, the Raven
coefficient is negative and statistically significant ($-0.164$, $p=0.051$). The Raven$\times$
mechanism interactions are imprecisely estimated and none is statistically significant at
conventional levels. Thus, we do not find robust evidence that the Raven-truth relationship differs
systematically across mechanisms: higher Raven scores are associated with a higher likelihood of
deviating from the induced ranking across all treatments.

The pooled rank regression (column 3) shows no significant relationship between cognitive ability
and assigned rank. Only RM improves ranks relative to DA in this pooled specification. The interaction model
(column 4) shows that the Raven-rank relationship differs across mechanisms. Under DA, higher Raven scores
predict modestly better outcomes (lower ranks): $-0.073$ ($p=0.027$). Under EADA, the interaction is positive
and significant ($0.144$, $p=0.008$), so the net association reverses in point estimate: $-0.073 + 0.144 = +0.071$,
consistent with higher-Raven subjects achieving worse outcomes under EADA. Under RM, the interaction is small and
statistically insignificant ($-0.041$), so the Raven-rank relationship is not statistically distinguishable from DA.

The EADA pattern is consistent with a sophistication trap: higher-ability subjects deviate more (as the truth-telling
regressions indicate), yet under EADA higher Raven scores are associated with worse assignments, while under DA
misreporting cannot improve one’s assignment relative to truth-telling. 
When the full CDE ability measure is used instead (Appendix~\ref{app:cde}), the Raven$\times$EADA interaction remains positive ($0.090$) but no longer reaches conventional significance ($p=0.117$). The direction of the sophistication trap is thus robust to the ability measure, though its statistical precision is not.
Taken together, these results do not support
Hypothesis~\ref{hyp:cognitive} in its intended form. Higher-ability subjects are not more likely to report truthfully under
any mechanism; if anything, the opposite holds. 

\begin{result}
	\label{res:cognitive}
Higher cognitive ability is associated with less truthful reporting across all mechanisms
($\hat\beta=-0.105$, $p=0.035$ with block C; $\hat\beta=-0.056$, $p=0.078$ with CDE), contrary to Hypothesis~\ref{hyp:cognitive}(i), and we find no evidence that this relationship
differs across treatments. 

In outcome regressions, {the Raven$\times$EADA interaction is positive in both specifications ($0.144$, $p=0.008$ with block C scores; $0.090$, $p=0.117$ with full CDE scores), consistent with a sophistication trap under EADA in which higher-ability subjects achieve worse outcomes.} Under RM, the Raven$\times$RM
interaction is statistically insignificant {\ in both specifications}.
\end{result}

Beyond this result, in Appendix~\ref{app:further} we examine whether these ability-based differences in behavior translate into ability-based sorting across schools. 
We find that DA produces the least sorting on both measures, and that EADA generates significantly more sorting than DA on one measure, consistent with the sophistication trap displacing higher-ability subjects to worse schools. 
However, most pairwise comparisons do not reach significance at conventional levels.

\section{Conclusion}
\label{sec:conclusion}

We conduct the first school choice experiment with parents, comparing the behavior and outcomes generated by the celebrated Deferred Acceptance and two more recent and manipulable mechanisms: Efficiency Adjusted Deferred Acceptance and Rank Minimizing. We find that parents and students deviate from their induced preferences at similar rates across mechanisms, yet the nature of deviations differs: students' deviations are more consistent with avoidable mistakes, whereas parents' deviations appear more deliberate. We also find that all mechanisms are frequently manipulated, but the consequences of manipulations vary sharply. Under DA, manipulations are rarely consequential, and when they are, they always harm the manipulator. Under EADA and RM, manipulations are consequential much more often; under EADA they are always harmful, while under RM a small fraction succeed. Cognitively sophisticated agents are more likely to misreport across all treatments, and under EADA the point estimates suggest these deviations lead to worse outcomes, consistent with a sophistication trap.

Despite the widespread manipulation of all mechanisms, the theoretical ranking of mechanisms by the traditional desiderata of efficiency and stability predicted under truthful behavior largely survives. RM is more rank- and Pareto-efficient but generates significantly more justified envy than DA, which in turn is never Pareto efficient and exhibits a poor rank distribution of its placements. We do not find evidence that EADA improves efficiency relative to DA, but it increases justified envy to twice DA levels.

Our results suggest that DA protects families through two distinct channels. Strategy-proofness ensures that no deviation can improve one's assignment (a property that both EADA and RM fail in theory, and that RM fails in practice). But DA also exhibits a second property in our data: most deviations do not alter the assignment at all. For policymakers seeking to level the playing field (and willing to tolerate an inefficient placement of students), DA remains the strongest choice: not because families report truthfully, which does not happen in practice, but because the mechanism ensures that failing to do so carries little consequence. Whether DA’s robustness to typical deviations can be established as a general theoretical property remains an open question.

Our experiment opens a new line of research within the large literature on strategic behavior in allocation mechanisms. The finding that parents and students behave similarly in aggregate, but differ in the nature of their mistakes, has implications for how policymakers interpret existing experimental evidence. Future work can extend this comparison to other mechanisms, information structures, and parent populations, and test whether interventions that reduce manipulation among students are equally effective for parents.
\bigskip

\singlespacing 
\setlength{\parskip}{-0.2em} 


	\bibliographystyle{ACM-Reference-Format}
\bibliography{bibliogr}

\newpage
\appendix
\setcounter{secnumdepth}{3}
\section{Preferences and Priorities Used}
\label{app:preferencesused}

As described in the main text, our experiment uses a school choice problem with $18$ subjects competing for $18$ seats distributed across $7$ schools.
Schools $s_1$ and $s_2$ each have capacity $q_{s_1}=q_{s_2}=4$, while schools $s_3$ through $s_7$ each have capacity $q_s=2$.

The preferences we used in the experiment appear in Table \ref{tab:preferences} below.
\begin{table}[H]
\centering
\caption{Preferences used in the experiment.}
\label{tab:preferences}
\resizebox{\textwidth}{!}{%
	\begin{tabular}{*{18}{c}}
		\toprule
		$i_1$ & $i_2$ & $i_3$ & $i_4$ & $i_5$ & $i_6$ & $i_7$ & $i_8$ & $i_9$ & $i_{10}$ & $i_{11}$ & $i_{12}$ & $i_{13}$ & $i_{14}$ & $i_{15}$ & $i_{16}$ & $i_{17}$ & $i_{18}$ \\
		\midrule
		$s_1$ & $s_2$ & $s_1$ & $s_2$ & $s_1$ & $s_2$ & $s_2$ & $s_2$ & $s_1$ & $s_2$  & $s_2$  & $s_2$  & $s_5$  & $s_2$  & $s_1$  & $s_2$  & $s_7$  & $s_2$  \\
		$s_2$ & $s_4$ & $s_3$ & $s_1$ & $s_2$ & $s_3$ & $s_1$ & $s_7$ & $s_2$ & $s_5$  & $s_7$  & $s_4$  & $s_1$  & $s_5$  & $s_5$  & $s_5$  & $s_4$  & $s_7$  \\
		$s_6$ & $s_7$ & $s_7$ & $s_7$ & $s_3$ & $s_5$ & $s_3$ & $s_3$ & $s_5$ & $s_1$  & $s_3$  & $s_1$  & $s_7$  & $s_1$  & $s_7$  & $s_1$  & $s_1$  & $s_3$  \\
		$s_4$ & $s_1$ & $s_2$ & $s_3$ & $s_6$ & $s_1$ & $s_5$ & $s_5$ & $s_4$ & $s_6$  & $s_1$  & $s_6$  & $s_2$  & $s_4$  & $s_6$  & $s_6$  & $s_5$  & $s_6$  \\
		$s_7$ & $s_5$ & $s_5$ & $s_6$ & $s_5$ & $s_4$ & $s_7$ & $s_1$ & $s_3$ & $s_3$  & $s_4$  & $s_5$  & $s_3$  & $s_7$  & $s_2$  & $s_3$  & $s_2$  & $s_1$  \\
		$s_3$ & $s_3$ & $s_4$ & $s_5$ & $s_4$ & $s_6$ & $s_6$ & $s_6$ & $s_7$ & $s_7$  & $s_5$  & $s_7$  & $s_4$  & $s_3$  & $s_3$  & $s_7$  & $s_3$  & $s_4$  \\
		$s_5$ & $s_6$ & $s_6$ & $s_4$ & $s_7$ & $s_7$ & $s_4$ & $s_4$ & $s_6$ & $s_4$  & $s_6$  & $s_3$  & $s_6$  & $s_6$  & $s_4$  & $s_4$  & $s_6$  & $s_5$  \\
		\bottomrule
	\end{tabular}%
}
\end{table}

The priorities used appear in Table \ref{tab:priorities1} below.
\begin{table}[H]
\centering
\caption{Priorities used in the experiment.}
\label{tab:priorities1}
\begin{tabular}{*{7}{c}}
	\toprule
	$s_1$ & $s_2$ & $s_3$ & $s_4$ & $s_5$ & $s_6$ & $s_7$ \\
	\midrule
	$i_{10}$ & $i_1$  & $i_2$  & $i_{17}$ & $i_{16}$ & $i_{17}$ & $i_4$  \\
	$i_4$   & $i_6$  & $i_1$  & $i_{18}$ & $i_7$  & $i_8$  & $i_{13}$ \\
	$i_{15}$ & $i_{12}$ & $i_9$  & $i_6$  & $i_{12}$ & $i_5$  & $i_9$  \\
	$i_1$   & $i_{15}$ & $i_{10}$ & $i_2$  & $i_{10}$ & $i_2$  & $i_7$  \\
	$i_6$   & $i_{18}$ & $i_5$  & $i_1$  & $i_5$  & $i_4$  & $i_{10}$ \\
	$i_{14}$ & $i_4$  & $i_3$  & $i_5$  & $i_{13}$ & $i_9$  & $i_8$  \\
	$i_{17}$ & $i_9$  & $i_6$  & $i_{13}$ & $i_{14}$ & $i_{16}$ & $i_{18}$ \\
	$i_{13}$ & $i_8$  & $i_{15}$ & $i_{16}$ & $i_6$  & $i_{18}$ & $i_1$  \\
	$i_5$   & $i_{13}$ & $i_{16}$ & $i_{11}$ & $i_{17}$ & $i_{12}$ & $i_2$  \\
	$i_{11}$ & $i_{14}$ & $i_{11}$ & $i_7$  & $i_2$  & $i_{11}$ & $i_6$  \\
	$i_{12}$ & $i_{10}$ & $i_{14}$ & $i_9$  & $i_{15}$ & $i_3$  & $i_{12}$ \\
	$i_8$   & $i_{16}$ & $i_{18}$ & $i_{12}$ & $i_9$  & $i_6$  & $i_{15}$ \\
	$i_3$   & $i_5$  & $i_4$  & $i_8$  & $i_4$  & $i_1$  & $i_{16}$ \\
	$i_7$   & $i_7$  & $i_8$  & $i_{10}$ & $i_1$  & $i_{10}$ & $i_{17}$ \\
	$i_{16}$ & $i_3$  & $i_{17}$ & $i_3$  & $i_3$  & $i_7$  & $i_5$  \\
	$i_9$   & $i_{17}$ & $i_7$  & $i_{14}$ & $i_8$  & $i_{14}$ & $i_3$  \\
	$i_{18}$ & $i_{11}$ & $i_{12}$ & $i_4$  & $i_{18}$ & $i_{15}$ & $i_{14}$ \\
	$i_2$   & $i_2$  & $i_{13}$ & $i_{15}$ & $i_{11}$ & $i_{13}$ & $i_{11}$ \\
	\bottomrule
\end{tabular}
\end{table}

With these preferences and priorities, we obtain the following matchings with our three mechanisms of interest:
\begin{align*}
\mu^{DA} &=(s_1\text{:}\{i_1,i_{10},i_{14},i_{15}\},\;s_2\text{:}\{i_4,i_6,i_{12},i_{18}\},\;s_3\text{:}\{i_5,i_9\},\\
& s_4\text{:}\{i_2,i_{17}\},\;s_5\text{:}\{i_7,i_{16}\},\;s_6\text{:}\{i_3,i_{11}\},\;s_7\text{:}\{i_8,i_{13}\}),\\ 
\mu^{EADA} &=(s_1\text{:}\{i_1,i_{13},i_{14},i_{15}\},\;s_2\text{:}\{i_4,i_6,i_{12},i_{18}\},\;s_3\text{:}\{i_5,i_7\},\\
& s_4\text{:}\{i_2,i_9\},\;s_5\text{:}\{i_{10},i_{16}\},\;s_6\text{:}\{i_3,i_{11}\},\;s_7\text{:}\{i_8,i_{17}\})\\
\mu^{RM} &=(s_1\text{:}\{i_4,i_5,i_9,i_{15}\},\;s_2\text{:}\{i_7,i_{11},i_{14},i_{18}\},\;s_3\text{:}\{i_3,i_6\},\\
&\hspace{1.6em} s_4\text{:}\{i_2,i_{12}\},\;s_5\text{:}\{i_{13},i_{16}\},\;s_6\text{:}\{i_1,i_{10}\},\;s_7\text{:}\{i_8,i_{17}\}.
\end{align*}

The matchings generated have the following properties, summarized in Table \ref{tab:properties1}.
\begin{table}[H]\centering
\small

\begin{tabular}{lcccccc}
	\toprule
	Mechanism & PE? & JE? & \#JE triples & \#JE students & Rank profile $\rho$ & Avg rank $\bar r$ \\
	\midrule
	DA   & No  & No  & $0$  & $0$ & $(6,4,4,1,1,0,2)$ & $49/18$ \\
	EADA & Yes & Yes & $3$  & $2$ & $(7,5,3,1,0,0,2)$ & $44/18$ \\
	RM   & Yes & Yes & $30$ & $8$ & $(9,7,1,1,0,0,0)$ & $30/18$ \\
	\bottomrule
\end{tabular}
\caption{PE and JE refer to Pareto-efficient and justified envy. A justified-envy (JE) triple is $(i,s,j)$ with $i$ preferring $s$ to $\mu(i)$ and having higher priority at $s$ than assignee $j\in\mu(s)$. The rank profile $\rho(\mu)=(\rho_1,\ldots,\rho_7)$ reports a histogram of outcomes, where $\rho_k$ denotes the number of students who are assigned their $k$-th ranked school under matching $\mu$ (so DA matches 6 subjects to their first-ranked school, 4 to the second, and so on).}
\label{tab:properties1}
\end{table}

\newpage
\section{Experimental Instructions}
\label{appendix}
Here we provide verbatim the experimental instructions provided to participants as well as the quiz they were presented with.

\subsection{Instructions}

\medskip

{
\setlength{\parindent}{0pt}

\begin{center}
\textbf{Welcome}
\end{center}

\medskip

This experiment will take around 45 minutes. It is divided into two parts. Please give it your full attention.

\medskip

In the first part, you will be asked to submit a ranking of potential schools. An algorithm will use the rankings of schools reported by you and other participants to assign you to one school. You will first be given an explanation of the algorithm. Then, you will be asked a few questions to ensure that you have understood the procedure. Finally, you will be asked to rank seven schools. You will be paid up to \pounds 55 based on your decisions and the decisions of other parents in this first part of the experiment.

\medskip

In the second part of the experiment, you will be asked to complete an IQ test in which you will have to identify matching patterns. You can receive up to an additional \pounds 3.6 based on your performance.

\medskip

In addition, every participant will receive \pounds 5 for their attendance. Therefore, you may earn up to \pounds 63.6 in total. 

\medskip

Phones may not be used during the experiment, and you may not communicate with other parents. If you need assistance at any point, please raise your hand.

\medskip

This study is conducted by academics from the University of Essex, Queen’s University of Belfast, and the Centre for European Economic Research. Participation in this study is anonymous, and the study has received ethical approval.

}

\medskip

{
\setlength{\parindent}{0pt}

\begin{center}
\textbf{How to Rank Schools}
\end{center}

\medskip

There are seven schools with a total of 18 available seats. All parents in the experiment (18 parents in total) will be asked to rank these schools privately. An algorithm will then be used to assign parents to schools. The algorithm will be explained to you next.

\medskip

You will be paid depending on which school you are assigned to. The school that gives other parents a higher payment may be different from the school that gives you a higher payment.

\medskip

The school to which you are assigned depends on the ranking of schools that you submit. You can change the ranking of schools by placing them into different positions. Please practice how to rank schools below. You cannot continue with the experiment until you have ranked all schools.

\medskip

\begin{figure}[H]
    \centering
    \includegraphics[width=1\textwidth]{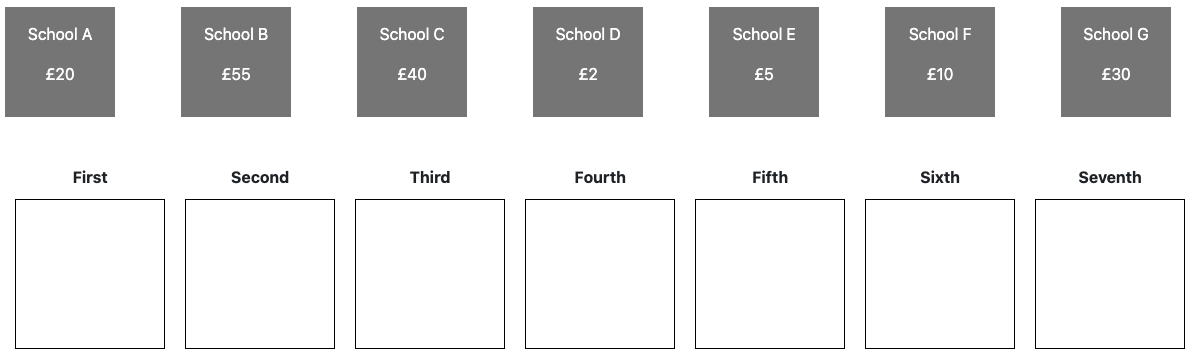}
\end{figure}
}

\medskip

\setlength{\parindent}{0pt}

\begin{center}
\textbf{Description of the algorithm}
\end{center}

\medskip

\begin{center}
\textbf{Deferred Acceptance (DA)}
\end{center}

\medskip

The school seats are allocated using the following procedure. Read this explanation in detail, as misunderstandings can significantly affect your payment.

\medskip

\textbf{Step 1:} You and all other parents will rank all schools privately. 

\medskip

\textbf{Step 2:} We will send your application to the school you have ranked first. If a school receives more applications than the number of seats available, it temporarily accepts parents with the highest priority, one at a time, up to its capacity. The remaining applicants are permanently rejected.

\medskip

\textbf{Step 3:} If your application was rejected at your first-ranked school, we will send your application to your second-best ranked one. If this school has empty seats, it will temporarily accept your application. If the school is full, it will compare your application to the other new applicants and those that it previously accepted. It temporarily accepts the ones with the highest priority, one at a time, up to its capacity and permanently rejects the remaining ones.

\medskip

\textbf{Step 4:} If your application was rejected in the previous step, we will send your application to the third-best ranked school, where it will again be compared against new applicants and those temporarily accepted. If your application is rejected again, we will send it to the fourth-best ranked school, and so on, until we eventually send it to the seventh-best ranked school if your application has been rejected by all other six schools.

\medskip

\textbf{Step 5:} The procedure stops, and every parent is assigned to a school.

\medskip

Now you will be guided through an example and asked a question to make sure you understood the procedure. Once you answer this question correctly, you will be allowed to rank schools.

\medskip

\textbf{Example}

There are three students, called Ann, Bob, and Carl, and three schools, with one seat available each, named School A, School B, and School C. Students rank schools as follows:

\medskip

\begin{center}
\begin{tabular}{|c|c|c|c|}
\hline
 & First Choice & Second Choice & Third Choice \\
\hline
Ann & School B & School A & School C \\
Bob & School A & School B & School C \\
Carl & School A & School C & School B \\
\hline
\end{tabular}
\end{center}

\medskip

This table means that Ann prefers School B, then School A, then School C. The schools have the following priorities:

\medskip

\begin{center}
\begin{tabular}{|c|c|c|c|}
\hline
 & First Priority & Second Priority & Third Priority \\
\hline
School A & Ann & Carl & Bob \\
School B & Bob & Ann & Carl \\
School C & Carl & Ann & Bob \\
\hline
\end{tabular}
\end{center}

\medskip

This second table means that Bob has the highest priority at School B, then Ann, then Carl. Therefore, if School B had to decide between the three of them, it would choose Bob. Applying the allocation procedure described above:

\medskip

\textbf{Note:} In each round, any student temporarily admitted at a school appears in framed text.

\medskip

\textbf{Round 1:} Ann applies to School B. Bob and Carl apply to School A.  

\begin{itemize}
    \item School A compares Bob and Carl. Bob has a lower priority, so he is rejected.
    \item Ann and Carl are temporarily accepted to Schools B and A.
\end{itemize}

\begin{center}
\begin{tabular}{|c|c|c|}
\hline
School A & School B & School C \\
\hline
Bob, \fbox{Carl} & \fbox{Ann} &  \\
\hline
\end{tabular}
\end{center}

\medskip

\textbf{Round 2:} Bob applies to School B.  

\begin{itemize}
    \item School B compares Ann and Bob. Ann has a lower priority, so she is rejected.
    \item Carl and Bob are temporarily accepted to Schools A and B.
\end{itemize}

\begin{center}
\begin{tabular}{|c|c|c|}
\hline
School A & School B & School C \\
\hline
\fbox{Carl} & Ann, \fbox{Bob} &  \\
\hline
\end{tabular}
\end{center}

\medskip

\textbf{Round 3:} Ann applies to School A.  

\begin{itemize}
    \item School A compares Ann to Carl. Carl has a lower priority, so he is rejected.
    \item Ann and Bob are temporarily accepted to Schools A and B.
\end{itemize}

\begin{center}
\begin{tabular}{|c|c|c|}
\hline
School A & School B & School C \\
\hline
Carl, \fbox{Ann} & \fbox{Bob} &  \\
\hline
\end{tabular}
\end{center}

\medskip

\textbf{Round 4:} Carl applies to School C.  

\begin{itemize}
    \item All schools have now filled their seats. The algorithm ends.
\end{itemize}

\begin{center}
\begin{tabular}{|c|c|c|}
\hline
School A & School B & School C \\
\hline
\fbox{Ann} & \fbox{Bob} & \fbox{Carl} \\
\hline
\end{tabular}
\end{center}

\medskip

The final allocation appears in framed text below:

\begin{center}
\begin{tabular}{|c|c|c|c|}
\hline
 & First Choice & Second Choice & Third Choice \\
\hline
Ann & School B & \fbox{School A} & School C \\
Bob & School A & \fbox{School B} & School C \\
Carl & School A & \fbox{School C} & School B \\
\hline
\end{tabular}
\end{center}

\medskip

This is, Ann is assigned to School A, Bob to School B, and Carl to School C.

\medskip

\textbf{Control question:} Consider a different example with 3 students (Peter, Mary, and Tom) and 3 schools (X, Y, Z). If in Round 1, two students Peter and Mary apply to School X, and Peter has a higher priority than Mary at this school, does it mean that Peter will be allocated to School X in the final allocation?

\setlength{\parindent}{0pt}

\medskip

\begin{center}
\textbf{Efficiency Adjusted Deferred Acceptance (EADA)}
\end{center}

\medskip

The school seats are allocated using the following procedure. Read this explanation in detail, as misunderstandings can significantly affect your payment.

\medskip

\textbf{Phase 1}

\medskip

\textbf{Step 1:} You and all other parents will rank all schools privately. 

\medskip

\textbf{Step 2:} We will send your application to the school you have ranked first. If a school receives more applications than the number of seats available, it temporarily accepts parents with the highest priority, one at a time, up to its capacity. The remaining applicants are permanently rejected.

\medskip

\textbf{Step 3:} If your application was rejected at your first-ranked school, we will send your application to your second-best ranked one. If this school has empty seats, it will temporarily accept your application. If the school is full, it will compare your application to the other new applicants and those that it previously accepted. It temporarily accepts the ones with the highest priority, one at a time, up to its capacity and permanently rejects the remaining ones.

\medskip

\textbf{Step 4:} If your application was rejected in the previous step, we will send your application to the third-best ranked school, where it will again be compared against new applicants and those temporarily accepted. If your application is rejected again, we will send it to the fourth-best ranked school, and so on, until we eventually send it to the seventh-best ranked school if your application has been rejected by all other six schools.

\medskip

\textbf{Step 5:} The procedure stops, and every parent is assigned to a school.

\medskip

\textbf{Phase 2}

\medskip

\textbf{Step 1:} Once the temporary allocation in Phase 1 has been obtained, the computer will identify the schools that are under-demanded: these are schools that no parent prefers to their own assigned school, or in other words, schools that did not reject any student in Phase 1. We will permanently match these schools to the parents assigned to them, and remove those schools and parents from the problem.

\medskip

\textbf{Step 2:} For the remaining parents and schools, we will re-do the procedure in Phase 1. No parent will be in a worse position than in Phase 1, because the algorithm will assign all parents to either the same school as in Phase 1, or to a more highly ranked school.

\medskip

\textbf{Step 3:} We will continue identifying under-demanded schools, i.e., those who did not reject anyone. These schools will permanently be assigned to the parents assigned to them, and the Phase 1 procedure will be applied to the remaining parents and schools, until all parents are permanently assigned to a school.

\medskip

Note that in each iteration of Phase 2, parents who have not been matched to an under-demanded school are either assigned to the same school as before, or to a more highly ranked school.

\medskip

Now you will be guided through an example and asked a question to make sure you understood the procedure. Once you answer this question correctly, you will be allowed to rank schools.

\medskip

\textbf{Example}

There are three students, called Ann, Bob, and Carl, and three schools, with one seat available each, named School A, School B, and School C. Students rank schools as follows:

\medskip

\begin{center}
\begin{tabular}{|c|c|c|c|}
\hline
 & First Choice & Second Choice & Third Choice \\
\hline
Ann & School B & School A & School C \\
Bob & School A & School B & School C \\
Carl & School A & School C & School B \\
\hline
\end{tabular}
\end{center}

\medskip

This table means that Ann prefers School B, then School A, then School C. The schools have the following priorities:

\medskip

\begin{center}
\begin{tabular}{|c|c|c|c|}
\hline
 & First Priority & Second Priority & Third Priority \\
\hline
School A & Ann & Carl & Bob \\
School B & Bob & Ann & Carl \\
School C & Carl & Ann & Bob \\
\hline
\end{tabular}
\end{center}

\medskip

\textbf{Round 1 (Phase 1):} Ann applies to School B. Bob and Carl apply to School A.  

\begin{itemize}
    \item School A compares Bob and Carl. Bob has a lower priority, so he is rejected.
    \item Ann and Carl are temporarily accepted to Schools B and A.
\end{itemize}

\begin{center}
\begin{tabular}{|c|c|c|}
\hline
School A & School B & School C \\
\hline
Bob, \fbox{Carl} & \fbox{Ann} &  \\
\hline
\end{tabular}
\end{center}

\medskip

\textbf{Round 2 (Phase 1):} Bob applies to School B.  

\begin{itemize}
    \item School B compares Ann and Bob. Ann has a lower priority, so she is rejected.
    \item Carl and Bob are temporarily accepted to Schools A and B.
\end{itemize}

\begin{center}
\begin{tabular}{|c|c|c|}
\hline
School A & School B & School C \\
\hline
\fbox{Carl} & Ann, \fbox{Bob} &  \\
\hline
\end{tabular}
\end{center}

\medskip

\textbf{Round 3 (Phase 1):} Ann applies to School A.  

\begin{itemize}
    \item School A compares Ann to Carl. Carl has a lower priority, so he is rejected.
    \item Ann and Bob are temporarily accepted to Schools A and B.
\end{itemize}

\begin{center}
\begin{tabular}{|c|c|c|}
\hline
School A & School B & School C \\
\hline
Carl, \fbox{Ann} & \fbox{Bob} &  \\
\hline
\end{tabular}
\end{center}

\medskip

\textbf{Round 4 (Phase 1):} Carl applies to School C.  

\begin{itemize}
    \item All schools have now filled their seats. The preliminary allocation is: Ann attends School A, Bob attends School B, Carl attends School C.
\end{itemize}

\begin{center}
\begin{tabular}{|c|c|c|}
\hline
School A & School B & School C \\
\hline
\fbox{Ann} & \fbox{Bob} & \fbox{Carl} \\
\hline
\end{tabular}
\end{center}

\medskip

\textbf{Phase 2 begins:} We find the school who rejected nobody, which is School C, and assign Carl. Remove School C and Carl from the problem. The simplified problem looks as follows:

\medskip

\begin{center}
\begin{tabular}{|c|c|c|}
\hline
 & First Choice & Second Choice \\
\hline
Ann & School B & School A \\
Bob & School A & School B \\
\hline
\end{tabular}
\end{center}

\medskip

\begin{center}
\begin{tabular}{|c|c|c|}
\hline
 & First Priority & Second Priority \\
\hline
School A & Ann & Bob \\
School B & Bob & Ann \\
\hline
\end{tabular}
\end{center}

\medskip

\textbf{Round 1 (Phase 2):} Ann applies to School B. Bob applies to School A.  

\begin{itemize}
    \item All remaining schools have now filled their seats. The algorithm ends.
\end{itemize}

\begin{center}
\begin{tabular}{|c|c|}
\hline
School A & School B \\
\hline
\fbox{Bob} & \fbox{Ann} \\
\hline
\end{tabular}
\end{center}

\medskip

The final allocation appears in framed text below:

\begin{center}
\begin{tabular}{|c|c|c|c|}
\hline
 & First Choice & Second Choice & Third Choice \\
\hline
Ann & \fbox{School B} & School A & School C \\
Bob & \fbox{School A} & School B & School C \\
Carl & School A & \fbox{School C} & School B \\
\hline
\end{tabular}
\end{center}

\medskip

This is, Ann is assigned to School B, Bob to School A, and Carl to School C.

\medskip

\textbf{Control question:} If students' 
preferences are given by the following Table, is there a school that rejects nobody in Phase 1?

\medskip

\begin{center}
\begin{tabular}{|c|c|c|c|}
\hline
 & First Choice & Second Choice & Third Choice \\
\hline
Ann & School A & School B & School C \\
Bob & School A & School B & School C \\
Carl & School A & School B & School C \\
\hline
\end{tabular}
\end{center}

\setlength{\parindent}{0pt}

\medskip

\pagebreak

\begin{center}
\textbf{Rank-Minimizing (RM)}
\end{center}

\medskip

The school seats are allocated using the following procedure. Read this explanation in detail, as misunderstandings can significantly affect your payment.

\medskip

\textbf{Step 1:} You and all other parents will rank all schools privately. 

\medskip

\textbf{Step 2:} If you are assigned to a school that you ranked first, we say that you have a rank equal to 1. Similarly, if you are assigned to the school that you ranked second, you have a rank equal to 2. If you are assigned to the school which you ranked last (remember, there are 7 schools), we say that you have a rank of 7.

\medskip

\textbf{Step 3:} After all parents have ranked all schools, a computer will find an allocation that has the lowest sum of ranks for all parents. In other words, the computer will find an assignment that tries to give everyone as highly-ranked a school as possible. If there are many such assignments, the computer will select one randomly.

\medskip

Now you will be guided through an example and asked a question to make sure you understood the procedure. Once you answer this question correctly, you will be allowed to rank schools.

\medskip

\textbf{Example}

There are three students, called Ann, Bob, and Carl, and three schools, with one seat available each, named School A, School B, and School C. Students rank schools as follows:

\medskip

\begin{center}
\begin{tabular}{|c|c|c|c|}
\hline
 & First Choice & Second Choice & Third Choice \\
\hline
Ann & School B & School A & School C \\
Bob & School A & School B & School C \\
Carl & School A & School C & School B \\
\hline
\end{tabular}
\end{center}

\medskip

This table means that Ann prefers School B, then School A, then School C. The schools have the following priorities:

\medskip

\begin{center}
\begin{tabular}{|c|c|c|c|}
\hline
 & First Priority & Second Priority & Third Priority \\
\hline
School A & Ann & Carl & Bob \\
School B & Bob & Ann & Carl \\
School C & Carl & Ann & Bob \\
\hline
\end{tabular}
\end{center}

\medskip

This second table means that Bob has the highest priority at School B, then Ann, then Carl. However, note that school priorities are irrelevant for the rank-minimizing algorithm.

\medskip

\textbf{Example allocation:}  

If we allocated Ann to School C, Bob to School B, and Carl to School A:

\begin{itemize}
    \item Ann would get her third choice (rank 3),
    \item Bob would get his second choice (rank 2),
    \item Carl would get his first choice (rank 1).
\end{itemize}

The sum of ranks is \(3+2+1=6\).

\medskip

\textbf{Rank-minimizing allocation:}  

The allocation that achieves the minimum sum of ranks assigns:

\begin{itemize}
    \item Ann to School B (rank 1),
    \item Bob to School A (rank 1),
    \item Carl to School C (rank 2).
\end{itemize}

The sum of ranks is \(1+1+2=4\). This allocation is highlighted below using \fbox{}.

\medskip

\begin{center}
\begin{tabular}{|c|c|c|c|}
\hline
 & First Choice & Second Choice & Third Choice \\
\hline
Ann & \fbox{School B} & School A & School C \\
Bob & \fbox{School A} & School B & School C \\
Carl & School A & \fbox{School C} & School B \\
\hline
\end{tabular}
\end{center}

\medskip

\textbf{Control question:} What is the sum of ranks for an allocation that assigns Ann to School A, Bob to School B, and Carl to School C?

\medskip

\setlength{\parindent}{0pt}

\begin{center}
\textbf{Quiz}
\end{center}

\medskip

\begin{center}
\begin{tabularx}{\textwidth}{|X|X|X|X|X|X|X|}
\hline
\textbf{School A.} & \textbf{School B.} & \textbf{School F.} & \textbf{School D.} & \textbf{School G.} & \textbf{School C.} & \textbf{School E.} \\
\hline
\textbf{£55} & \textbf{£40} & \textbf{£30} & \textbf{£20} & \textbf{£10} & \textbf{£5} & \textbf{£2} \\
\hline
This school has \textbf{4} seats. & This school has \textbf{4} seats. & This school has \textbf{2} seats. & This school has \textbf{2} seats. & This school has \textbf{2} seats. & This school has \textbf{2} seats. & This school has \textbf{2} seats. \\
\hline
This school prioritizes \textbf{3} students over you. & This school prioritizes \textbf{0} students over you. & This school prioritizes \textbf{12} students over you. & This school prioritizes \textbf{4} students over you. & This school prioritizes \textbf{7} students over you. & This school prioritizes \textbf{1} student over you. & This school prioritizes \textbf{13} students over you. \\
\hline
\textbf{5} students would get their highest payment from this school. & \textbf{11} students would get their highest payment from this school. & \textbf{0} students would get their highest payment from this school. & \textbf{0} students would get their highest payment from this school. & \textbf{1} student would get their highest payment from this school. & \textbf{0} students would get their highest payment from this school. & \textbf{1} student would get their highest payment from this school. \\
\hline
\end{tabularx}
\end{center}

\medskip

\textbf{Questions:}

\begin{itemize}
    \item If you are assigned to School F, what is your payment? 
    \item If you are assigned to School A, what is your payment? 
    \item How many students does School C prioritize over you? 
    \item How many students does School G prioritize over you? 
    \item How many students would get their highest payment from attending School C?
    \item How many students would get their highest payment from attending School A? 
\end{itemize}

\medskip

Now consider the following situation:

\medskip

\begin{center}
\includegraphics[width=1.0\textwidth]{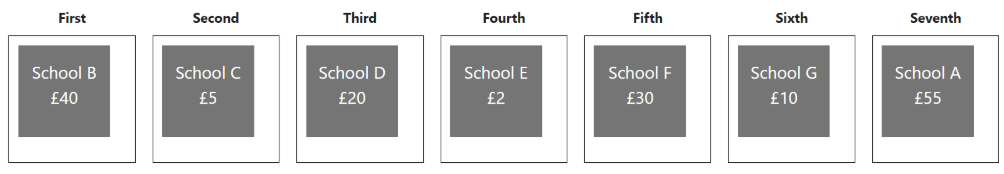}
\end{center}

\medskip

\begin{itemize}
    \item Which school is ranked third?  
    \item How much money is the player getting if they are assigned to their second choice?  
\end{itemize}

\medskip


\newpage
\section{Recombination Analysis}
\label{app:recomb}
This section assesses whether the comparative performance of the five mechanisms is robust to session composition in a controlled school-choice laboratory environment. The experiment features $K = 6$ independent lab sessions, each with $N = 18$ students and $7$ schools, with heterogeneous capacities summing to 18 seats. Students have induced (``true'') strict preferences over schools, described by a fixed payoff-based ordering (Table \ref{tab:preferences}), and they also submit stated preference lists that may differ from induced rankings. School priorities and capacities are held fixed throughout the recombination exercise.

\subsection{Recombination design}

We implement the recombinant estimation approach developed for games by \citet{mullin2006recombinant} and used in school-choice experiments following the recombination procedure described in \citet{calsamiglia2010constrained,calsamiglia2011comment}. Each recombination draw constructs a synthetic market by sampling, for each student $j$, that student’s preference from one of the donor sessions (with replacement). To preserve the dependence structure needed for valid inference, we iterate over all $(i,j)$ pairs -- session $i$ and student $j$ -- and, in each block, force student $j$'s preferences to come from session $i$ while sampling all other students’ columns from the donor pool. For each $(i,j)$, we generate $R = 10,000$ independent recombinations, yielding $K \times N \times R = 1{,}080{,}000$ synthetic markets per mechanism. In every synthetic market, we run the mechanism using the fixed priority structure and capacities.
The recombination exercise is designed to answer whether the welfare/stability conclusions are sensitive to the particular mix of types realized in a given lab session.

\subsection{Inference}

For each mechanism and outcome, we report the recombinant mean $\hat\mu$ and an asymptotic standard error using the Mullin–Reiley decomposition:
$$
\widehat{\mathrm{Var}}(\hat\mu) = \frac{\hat\sigma^2}{K N R}
+ \frac{N \hat\phi}{K},
$$
where $\hat\sigma^2$ is the variance of draw-level outcomes and $\hat\phi$ is a within-$(i,j)$ covariance component estimated via a half-split within each block. Because finite-sample noise can yield slightly negative $\hat\phi$, we truncate $\hat\phi$ at zero to enforce non-negativity of the variance estimator (a standard practical adjustment). Results are stable when increasing $R$ from 10,000 up to 200,000.

\bigskip

Overall, recombination confirms that the key comparative statics are not an artifact of any single lab session’s composition: RM robustly improves rank-based welfare and tail outcomes but at the cost of substantially higher justified envy, while DA-based mechanisms preserve stability but exhibit greater scope for Pareto-improving swaps under induced preferences.

\subsection{Truth-telling calibration in recombination}

In the baseline recombinant design, each student’s stated preference list in a synthetic market is drawn from the pool of donor sessions. In our data, the aggregate truth-telling rate across donor sessions is $p_d = 0.29$, where a student is classified as truth-telling if her submitted ranking coincides with her induced (true) ranking. To study mechanism performance under counterfactual reporting environments with different degrees of truthful reporting, we modify the recombination procedure so that the expected truth-telling rate in the synthetic markets equals a pre-specified target level $\tau \in \{0.30, 0.50, 0.70, 0.90\}$, while preserving the Mullin-Reiley $(i,j)$ block structure used for inference.

Concretely, in each recombinant draw we continue to (i) sample donor-session stated preferences with replacement and (ii) force the fixed student $j$ in block $(i,j)$ to be drawn from donor session $i$ as in the baseline procedure. For the remaining $N-1$ students in the synthetic market, we introduce an independent mixture rule: with probability $x(\tau)$ student $j'$ draws her submitted preference list from a randomly selected donor session (as in the baseline recombination), and with probability $1-x(\tau)$ we instead set her submitted list equal to her induced preference ordering (i.e., we impute truthful reporting for that student). Because truthful reports coincide with induced preferences by construction, the latter source has truth rate one.

We choose the mixing probability $x(\tau)$ so that the \emph{expected} truth-telling rate in an $N$-student synthetic market equals the desired target $\tau$. With the fixed student always donor-sourced, the expected truth rate is
\[
\tau \;=\; \frac{1}{N}p_d \;+\; \frac{N-1}{N}\Big( (1-x(\tau))\cdot 1 \;+\; x(\tau)\cdot p_d \Big),
\]
which yields the closed-form solution
\[
x(\tau) \;=\; \frac{p_d + (N-1) - N\tau}{(N-1)(1-p_d)}.
\]
Using $N=18$ and $p_d=0.29$, this implies the donor-draw shares
\[
x(0.30) \approx 0.993, \qquad
x(0.50) \approx 0.701, \qquad
x(0.70) \approx 0.389, \qquad
x(0.90) \approx 0.097,
\]
with the complementary shares $1-x(\tau)$ drawn from truthful (induced) preferences. Thus, higher targets $\tau$ correspond mechanically to replacing a larger fraction of donor-session submissions with induced rankings.

Feasibility requires $x(\tau)\in[0,1]$. Under our design (one fixed donor-sourced student per synthetic market), the attainable range of target truth rates is
\[
\tau \in \left[\frac{p_d}{N} + \frac{N-1}{N}p_d,\; \frac{p_d}{N} + \frac{N-1}{N}\cdot 1\right]
= \left[p_d,\; 1 - \frac{1-p_d}{N}\right].
\]
With $N=18$ and $p_d=0.29$, this corresponds to $\tau \in [0.29,\; 0.9606]$, so all targets in $\{0.30,0.50,0.70,0.90\}$ are feasible. This calibration changes only the source of submitted preferences for a subset of students while leaving priorities, capacities, the mechanism, and the recombination/inference structure unchanged. As a result, comparisons across mechanisms can be interpreted as robustness to alternative truth-telling counterfactuals, rather than to realized session composition alone.

\newpage

\section{EADA Step-by-Step Implementation}
\label{app:eadaex}
In the example, in the first DA run the only underdemanded school is $s_1$, which is matched to $i_3$;

\begin{table}[H]
\centering
\caption{DA proposal rounds (tentative holders shown; rejections in bold).}
\label{tab:rejections}
\begin{tabular}{ccccc}
	\toprule
	& $s_1$ & $s_2$ & $s_3$ & $s_4$ \\
	\midrule
	\text{Round 1} &  & $i_1,\pmb{i_2}$ &  & $\pmb{i_3}, i_4$ \\
	\text{Round 2} &  & $i_1$ & $i_3$ & $i_2, \pmb{i_4}$ \\
	\text{Round 3} &  & $i_1$ & $\pmb{i_3}, i_4$ & $i_2$ \\
	\text{Round 4} &  & $\pmb{i_1}, i_3$ & $i_4$ & $i_2$ \\
	\text{Round 5} &  & $i_3$ & $i_1, \pmb{i_4}$ & $i_2$ \\
	\text{Round 6} &  & $\pmb{i_3}, i_4$ & $i_1$ & $i_2$ \\
	\text{Round 7} & $i_3$ & $i_4$ & $i_1$ & $i_2$ \\
	\bottomrule
\end{tabular}
\end{table}

We remove $(i_3,s_1)$ and rerun DA on the reduced problem with students $\{i_1,i_2,i_4\}$ and schools $\{s_2,s_3,s_4\}$.

\begin{table}[H]
\caption{Second step in EADA: reduced preferences (left) and DA rounds (right).}
\label{tab:seada1}
\begin{minipage}{.52\linewidth}
	\centering
	\begin{tabular}{ccccccc}
		\toprule
		$i_1$ & $i_2$ & $i_4$ & &  $s_2$ & $s_3$ & $s_4$ \\
		\midrule
		$s_2$ & $s_2$ & $s_4$ & & $i_4$ & $i_2$ & $i_1$ \\
		$s_3$ & $s_4$ & $s_3$ & & $i_1$ & $i_4$ & $i_2$ \\
		$s_4$ & $s_3$ & $s_2$ & & $i_2$ & $i_1$ & $i_4$ \\
		\bottomrule
	\end{tabular}
\end{minipage}\hfill%
\begin{minipage}{.44\linewidth}
	\centering
	\begin{tabular}{cccc}
		\toprule
		& $s_2$ & $s_3$ & $s_4$ \\
		\midrule
		\text{Round 1} & $i_1,\pmb{i_2}$ &  & $i_4$ \\
		\text{Round 2} & $i_1$ &  & $i_2,\pmb{i_4}$ \\
		\text{Round 3} & $i_1$ & $i_4$ & $i_2$ \\
		\bottomrule
	\end{tabular}
\end{minipage}
\end{table}

In this reduced DA run, $s_3$ is underdemanded and is matched to $i_4$; we remove $(i_4,s_3)$ and rerun DA on the remaining problem with students $\{i_1,i_2\}$ and schools $\{s_2,s_4\}$.

\begin{table}[H]
\caption{Third step in EADA: reduced problem (left) and DA rounds (right).}
\label{tab:seada2}
\begin{minipage}{.52\linewidth}
	\centering
	\begin{tabular}{ccccc}
		\toprule
		$i_1$ & $i_2$ & & $s_2$ & $s_4$ \\
		\midrule
		$s_2$ & $s_4$ & & $i_1$ & $i_2$ \\
		$s_4$ & $s_2$ & & $i_2$ & $i_1$ \\
		\bottomrule
	\end{tabular}
\end{minipage}\hfill%
\begin{minipage}{.44\linewidth}
	\centering
	\begin{tabular}{ccc}
		\toprule
		& $s_2$ & $s_4$ \\
		\midrule
		\text{Round 1} & $i_1$ & $i_2$ \\
		\bottomrule
	\end{tabular}
\end{minipage}
\end{table}

In the final reduced instance, both schools are under-demanded and the algorithm terminates.
The resulting EADA matching is therefore
\[
\mu^{EADA} = (i_1\text{--}s_2,\; i_2\text{--}s_4,\; i_3\text{--}s_1,\; i_4\text{--}s_3),
\]
with rank profile $(1,2,4,2)$.

\newpage
\section{Further Analysis}
\label{app:further}

\paragraph{Where do subjects first deviate?}

Table~\ref{tab:firstdiff} reports the position at which subjects first deviate from their induced preferences, conditional on misreporting. Across treatments, roughly 60\% of misreporters deviate at the very first position, i.e.\ they do not rank their true first choice first. This behavior is consistent with not ranking schools that are deemed unattainable \citep{chen2019self}. Under DA, ranking one's true first choice first is always weakly optimal, yet most deviations begin at the top of the list. This pattern is consistent with heuristics such as ``do not waste a top slot on a competitive school'' or with mistaken beliefs about the returns to strategic positioning. Formal tests fail to reject equality of the first-deviation distributions across treatments ($\chi^2$ test of independence, $p=0.88$).

\begin{table}[h!]
	\centering
	\caption{Position of first deviation from true preferences (among those who misreport)}
	\label{tab:firstdiff}
	\begin{tabular}{lccccc}
		\toprule
		& DA & EADA & RM & sDA & sDAL \\
		\midrule
		Position 1 & 0.58 & 0.60 & 0.58 & 0.62 & 0.67 \\
		Position 2 & 0.29 & 0.22 & 0.25 & 0.17 & 0.21 \\
		Position 3 & 0.07 & 0.10 & 0.11 & 0.12 & 0.06 \\
		Position 4+ & 0.06 & 0.08 & 0.06 & 0.08 & 0.05 \\
		\bottomrule
	\end{tabular}
\end{table}

\paragraph{Skipping down and inflating demand.}

We document two recurring deviation patterns. To quantify them, we construct two indicators based on the submitted list relative to induced preferences. \emph{Skipping down} requires (i) at least one of the two high-demand schools (A/B) to be demoted relative to the true ranking, (ii) neither A nor B to be promoted, and (iii) the remaining five schools to preserve their true relative order. \emph{Inflating demand} requires (i) the true first choice to remain in position~1, (ii) at least one of A/B -- a high-demand school that is not one's true top choice -- to be promoted into position~2 or~3, and (iii) the remaining five schools to preserve their true relative order. Table~\ref{tab:skip_trunc} reports the prevalence of these deviations by treatment.

\begin{table}[h!]
	\centering
	\caption{Share of manipulations}
	\label{tab:skip_trunc}
	\begin{tabular}{lccccc}
		\toprule
		& \multicolumn{3}{c}{Parents} & \multicolumn{2}{c}{Students} \\
		\cmidrule(lr){2-4} \cmidrule(lr){5-6}
		& DA & EADA & RM & sDA & sDAL \\
		\midrule
		Share skipping down & 0.06 & 0.08 & 0.12 & 0.20 & 0.20 \\
		Share inflating demand  & 0.04 & 0.04 & 0.07 & 0.03 & 0.04 \\
		\bottomrule
	\end{tabular}
\end{table}

Skipping down is more common among students than parents, consistent with students making more impulsive deviations. The pattern reflects a ``do not waste a slot'' heuristic: subjects demote competitive schools they believe they cannot get into. Under DA, skipping down cannot be beneficial; at best it is inconsequential, and at worst it is harmful if an attainable school is moved too far down.

Inflating demand occurs at low rates across all mechanisms. Its strategic implications, however, differ. Under DA and EADA, promoting a popular school one does not truly prefer cannot improve one's assignment. Under RM, by contrast, inserting a high-demand school one does not truly prefer into position~2 or~3 inflates the apparent demand for that school in the rank-minimizing objective, making it costlier for the algorithm to assign other subjects to it, thereby increasing the likelihood that the manipulator retains her true first choice. Despite this, we do not observe inflating demand at meaningfully higher rates under RM than under other mechanisms.

The inflating demand pattern is consistent with the first-order heuristic strategy derived by \citet{tasnim2024strategic}, who show theoretically that agents can benefit under RM by keeping their true top choice fixed and promoting popular schools into subsequent positions. Our contribution is to document that real subjects in a laboratory setting, and without any strategic advice, independently arrive at behavior resembling this strategy, albeit at low rates. To the best of our knowledge, this is the first empirical observation of this manipulation pattern.

\paragraph{Ability sorting across schools.}

If cognitive ability influences strategic behavior differently across mechanisms, it may also produce different degrees of ability-based sorting across schools. \citet{basteck2018cognitive} document substantial ability segregation under the Boston mechanism relative to DA: low-ability students are overrepresented at less preferred schools, because higher-ability subjects successfully exploit the mechanism's obvious manipulability. Under DA, this segregation largely disappears. We examine whether analogous patterns arise when comparing DA to EADA and RM, using two measures computed via recombinant estimation. The \emph{between-school variance share} captures the fraction of total within-session variation in Raven scores attributable to differences in school-level average Raven scores; lower values indicate weaker sorting. The \emph{dispersion of school mean Raven} measures the standard deviation across schools of school-level average ability; lower values indicate schools are more similar in average cognitive ability. Table~\ref{tab:segregation} reports both measures.

\begin{table}[h!]
	\centering
	\caption{Ability sorting across schools (recombinant estimates)}
	\label{tab:segregation}
	\begin{tabular}{lccccc}
		\toprule
		& DA & EADA & RM & sDA & sDAL \\
		\midrule
		Between-school variance share & 0.320 & 0.359 & 0.368 & 0.339 & 0.395 \\
		& (0.026) & (0.026) & (0.023) & (0.029) & (0.023) \\[4pt]
		Dispersion of school mean Raven & 1.456 & 1.939 & 1.568 & 1.307 & 1.631 \\
		& (0.149) & (0.222) & (0.158) & (0.155) & (0.232) \\
		\bottomrule
		\multicolumn{6}{l}{\footnotesize Recombinant standard errors in parentheses.}
	\end{tabular}
\end{table}

DA produces the lowest ability sorting among parent treatments on both measures. On the between-school variance share, the differences are directionally consistent with strategy-proofness limiting ability-based sorting, but no pairwise comparison reaches significance (RM vs.\ DA: $p=0.084$; EADA vs.\ DA: $p=0.146$). On the dispersion of school mean Raven, EADA generates significantly more sorting than DA ($p=0.041$), while the remaining comparisons are not significant. This complements the findings of \citet{basteck2018cognitive} but through the opposite channel: in their setting, segregation arises because higher-ability subjects \emph{successfully} exploit the Boston mechanism's obvious manipulability and obtain better schools. In ours, the EADA pattern is driven by the sophistication trap documented in Section~\ref{subsec:ability}: higher-ability subjects attempt to exploit a mechanism whose profitable deviations are non-obvious, are harmed as a result, and are systematically displaced to worse schools. Both channels generate ability-based sorting, but for opposite reasons.

\section{Cognitive Ability: Alternative Specification}
\label{app:cde}

Table~\ref{tab:cognitive_cde} replicates the cognitive ability analysis from Section~\ref{subsec:ability} using the full Raven score (blocks C, D and E) instead of block C alone. This requires dropping treatment sDA, for which only block C answers were recorded, reducing the sample from 540 to 432 observations.

All results are directionally consistent with the main specification. In the pooled truth-telling logit (column 1), the Raven coefficient is negative and significant ($-0.135$, $p=0.002$), confirming that higher-ability subjects deviate more. In the interaction model (column 2), the Raven$\times$EADA coefficient on truth-telling is negative and marginally significant ($-0.228$, $p=0.063$), suggesting that higher-ability subjects deviate even more under EADA than under DA.

In the rank regressions, the Raven$\times$EADA interaction remains positive ($0.090$) but does not reach conventional significance ($p=0.117$), compared to $0.144$ ($p=0.008$) in the block C specification. The direction of the sophistication trap -- higher-ability subjects achieving worse outcomes under EADA -- is preserved, but estimated less precisely with the full ability measure.

\begin{table}[H]
	\centering
	\caption{Cognitive ability, truth-telling, and outcomes (blocks CDE)}
	\label{tab:cognitive_cde}
	\small
	\begin{tabular}{l cc cc}
		\toprule
		& \multicolumn{2}{c}{Truth-telling (logit)} & \multicolumn{2}{c}{Assigned rank (OLS)} \\
		\cmidrule(lr){2-3} \cmidrule(lr){4-5}
		& (1) & (2) & (3) & (4) \\
		\midrule
		Raven & $-0.135^{***}$ & $-0.110$ & $-0.070^{**}$ & $-0.056^{*}$ \\
		& $(0.044)$ & $(0.082)$ & $(0.028)$ & $(0.032)$ \\[4pt]
		EADA & $0.118$ & $1.838^{*}$ & $-0.079$ & $-0.794^{*}$ \\
		& $(0.310)$ & $(0.991)$ & $(0.156)$ & $(0.406)$ \\[4pt]
		RM & $-0.078$ & $-1.900^{*}$ & $-0.325^{*}$ & $0.285$ \\
		& $(0.321)$ & $(1.153)$ & $(0.178)$ & $(0.446)$ \\[4pt]
		sDAL & $-0.373$ & $-0.350$ & $0.164$ & $1.723^{*}$ \\
		& $(0.340)$ & $(1.323)$ & $(0.166)$ & $(0.956)$ \\[4pt]
		Raven $\times$ EADA & & $-0.228^{*}$ & & $0.090$ \\
		& & $(0.122)$ & & $(0.057)$ \\[4pt]
		Raven $\times$ RM & & $0.210$ & & $-0.070$ \\
		& & $(0.130)$ & & $(0.048)$ \\[4pt]
		Raven $\times$ sDAL & & $-0.005$ & & $-0.174^{*}$ \\
		& & $(0.152)$ & & $(0.100)$ \\[4pt]
		$\overline{\text{Raven}}_{-i}$ & & & $-0.014$ & $-0.032$ \\
		& & & $(0.083)$ & $(0.069)$ \\[4pt]
		Constant & $0.012$ & $-0.177$ & $3.498^{***}$ & $3.531^{***}$ \\
		& $(0.403)$ & $(0.676)$ & $(0.691)$ & $(0.561)$ \\
		\midrule
		$N$ & 432 & 432 & 432 & 432 \\
		\bottomrule
	\end{tabular}
	\vspace{4pt}
	
	{\footnotesize Notes: Raven denotes the total score on blocks C, D and E. Treatment sDA is excluded due to missing block D and E data. Columns (1)-(2) report logit coefficients with standard errors in parentheses; columns (3)-(4) report OLS estimates with standard errors clustered at the market level in parentheses.
		$^{*}p<0.10$, $^{**}p<0.05$, $^{***}p<0.01$. DA is the reference category. $\overline{\text{Raven}}_{-i}$ is included only in the rank regressions, as it should not affect one's truth-telling decision.}
\end{table}

The main differences between the two specifications are as follows. First, the pooled Raven-truth-telling relationship strengthens with CDE ($p=0.002$ versus $p=0.005$), so the finding that higher-ability subjects manipulate more is robust. Second, the Raven$\times$EADA interaction on assigned rank attenuates ($0.090$ versus $0.144$) and loses significance ($p=0.117$ versus $p=0.008$). Third, the Raven$\times$sDAL interaction on assigned rank becomes marginally significant ($-0.174$, $p=0.081$), suggesting that higher-ability students under low-stakes DA achieve better outcomes; this interaction is not significant in the block C specification. All other coefficients are qualitatively unchanged.

\end{document}